\documentclass[12pt]{article}
\usepackage{latexsym}
\usepackage{amsmath}
\usepackage{amssymb}
\usepackage{cite}

\date{}

\begin{document}
 \begin{center}
{\Large \textbf{Bi-Hamiltonian representation, symmetries and
integrals of mixed heavenly and Husain systems}\
\footnote{Preprint of an article accepted for publication in
Journal of Nonlinear Mathematical Physics, Copyright \copyright\
2010 by World Scientific Publishing Company, Journal URL:
http://www.worldscinet.com/jnmp/}}
 \vspace*{5mm} \\ {\large\textbf{M. B. Sheftel$^{1}$ and D.~Yaz{\i}c{\i}$^2$}}
\end{center}
 \vspace{2mm} $^1$ Department of Physics, Bo\u{g}azi\c{c}i
University 34342 Bebek, Istanbul, Turkey  \\
$^2$ Department of Physics, Y{\i}ld{\i}z Technical University,
34220 Esenler,\\ $\phantom{^2}$ Istanbul, Turkey
 \vspace{1mm}
 \\ E-mails: mikhail.sheftel@boun.edu.tr,
            yazici@yildiz.edu.tr

\begin{abstract}
\noindent In the recent paper by one of the authors (MBS) and A.
A. Malykh on the classification of second-order PDEs with four
independent variables that possess partner symmetries \cite{shma},
mixed heavenly equation and Husain equation appear as closely
related canonical equations admitting partner symmetries. Here for
the mixed heavenly equation and Husain equation, formulated in a
two-component form, we present recursion operators, Lax pairs of
Olver-Ibragimov-Shabat type and discover their Lagrangians,
symplectic and bi-Hamiltonian structure. We obtain all point and
second-order symmetries, integrals and bi-Hamiltonian
representations of these systems and their symmetry flows together
with infinite hierarchies of nonlocal higher symmetries.
\end{abstract}

\section{Introduction}
\setcounter{equation}{0}
 \label{sec-intro}

 In the recent paper \cite{shma}, one of the authors
(MBS) and A. A. Malykh obtained, up to a change of notation for
independent variables, the general form of second-order partial
differential equations (PDEs) with four independent variables
$t,x,y,z$, that possess partner symmetries
\cite{mns,mnsh,lift,shma08} and contain only second derivatives of
the unknown $u$:
\begin{gather}
 F = a_1(u_{ty}u_{xz} - u_{tz}u_{xy}) + a_2 (u_{tx}u_{ty} -
u_{tt}u_{xy})  + a_3(u_{ty}u_{xx} - u_{tx}u_{xy})\notag
\\ \mbox{} + a_4 (u_{tx}u_{tz} - u_{tt}u_{xz})
 + a_5 (u_{tz}u_{xx} - u_{tx}u_{xz}) + a_6(u_{tt}u_{xx} - u_{tx}^2)
 \label{Ffinal}
\\ \mbox{} + b_1 u_{xy} + b_2 u_{ty} + b_3 u_{xz} + b_4 u_{tz}
+ b_5 u_{tt} + 2b_6 u_{tx} + b_7 u_{xx} + b_0 = 0 ,
 \notag
\end{gather}
with constant coefficients $a_i$ and $b_i$. Partner symmetries,
that make it possible to obtain noninvariant solutions of PDEs of
the form (\ref{Ffinal}), are generated by the recursion relation:
\begin{eqnarray}
 & &\tilde{\varphi_t} =  - \big(a_2u_{ty} + a_4u_{tz} - a_6 u_{tx} + b_6
 - \omega_0\big)\varphi_t  -  \nonumber
 \\ & & \qquad\;\big(a_3u_{ty} + a_5u_{tz} + a_6 u_{tt} +
 b_7\big)\varphi_x + \big(a_1u_{tz} + a_2u_{tt} + a_3u_{tx} -
 b_1\big)\varphi_y \nonumber
 \\ & & \qquad\qquad\qquad\qquad\qquad \qquad\qquad\;\;
 \mbox{} + \big(- a_1u_{ty} + a_4u_{tt} + a_5u_{tx} -  b_3\big)\varphi_z ,
 \nonumber
 \\ & & \tilde{\varphi}_x = - \big(a_2u_{xy} + a_4u_{xz} - a_6 u_{xx} -
 b_5\big)\varphi_t -
  \label{recurs}
 \\ & &  \big(a_3u_{xy} + a_5u_{xz} + a_6 u_{tx} - b_6 - \omega_0\big)\varphi_x
   + \big(a_1u_{xz} + a_2u_{tx} + a_3u_{xx} + b_2\big)\varphi_y \nonumber
 \\ & & \qquad\qquad\qquad\qquad\qquad \qquad\qquad\;\;
 \mbox{} + \big(- a_1u_{xy} + a_4u_{tx} + a_5u_{xx} +  b_4\big)\varphi_z , \nonumber
 \end{eqnarray}
where $\varphi$ and $\tilde{\varphi}$ are symmetry characteristics
\cite{olv} and $\omega_0$ is a constant. In (\ref{Ffinal}) and
(\ref{recurs}), subscripts denote partial derivatives. The
transformation (\ref{recurs}) maps any symmetry $\varphi$ of the
equation (\ref{Ffinal}) into its partner symmetry
$\tilde{\varphi}$.

In \cite{shma}, we also listed canonical forms to which the
general form (\ref{Ffinal}) can be reduced by point and Legendre
transformations. Among these forms we find, along with the first
and second heavenly equations of Pleba\~nski \cite{pleb}, a new
equation that looks, up to a point, like the combination of these
two equations, which we called \textit{mixed heavenly equation:}
\begin{equation}\label{mix}
u_{ty}u_{xz} - u_{tz}u_{xy} + u_{tt}u_{xx}-u_{tx}^2 = \varepsilon,
\end{equation}
where $\varepsilon = \pm 1$. Recursion relation (\ref{recurs}) for
symmetries of the equation (\ref{mix}) becomes
\begin{gather}
  \tilde{\varphi}_t = (u_{tx} + \omega_0)\varphi_t -
u_{tt}\varphi_x + u_{tz}\varphi_y - u_{ty}\varphi_z ,\notag
 \\ \tilde{\varphi}_x = u_{xx}\varphi_t - (u_{tx} - \omega_0)\varphi_x + u_{xz}\varphi_y -
u_{xy}\varphi_z.
 \label{mixrecurs}
\end{gather}
Note that in our classification heavenly equations of Pleba\~nski
belong to equivalence classes different from the one to which the
mixed heavenly equation belongs, that is, they cannot be related
neither by point nor by Legendre transformations.\footnote{Quite
recently a different classification of integrable PDEs of
Pleba\~nski type was done in paper \cite{Fer}.}

Another form of a canonical equation from the same equivalence
class coincides, at $\varepsilon = + 1$, with the Husain heavenly
equation:
\begin{equation}\label{husain_u}
  u_{ty}u_{xz} - u_{tz}u_{xy} + u_{tt} + \varepsilon u_{xx} = 0 ,
\end{equation}
which is an alternative form of a basic self-dual gravity equation
arising in the chiral model approach to self-dual gravity
\cite{husain,ppg}. Recursion relation (\ref{recurs}) for
symmetries of equation (\ref{husain_u}) takes the form
\begin{gather}
\tilde{\varphi}_t = u_{tz}\varphi_y - u_{ty}\varphi_z -
\varepsilon \varphi_x + \omega_0\varphi_t ,\notag
\\ \tilde{\varphi}_x = u_{xz}\varphi_y -
u_{xy}\varphi_z + \varphi_t + \omega_0\varphi_x .
 \label{Husrecurs_u}
\end{gather}

Though equation (\ref{husain_u}) can be obtained from the mixed
heavenly equation (\ref{mix}) by Legendre transformation
(\ref{legmixhus}), the main objects of the Hamiltonian formulation
of Husain equation, like Lagrangian, symplectic two-form,
Hamiltonian operators and Hamiltonian densities cannot be obtained
that way. Therefore, we study Lax representation, symplectic and
Hamiltonian structures of Eq. (\ref{husain_u}) independently of
those of Eq. (\ref{mix}).

In this paper, we consider mixed heavenly equation and Husain
equation in a two-component form, which enables us to rewrite the
corresponding recursion relation as a single $2\times 2$
matrix-differential equation and introduce naturally recursion
operators $R$. Together with the operator $\hat{A}$ of the
symmetry condition, which determines symmetries in a two-component
form, the two operators $R$ and $\hat{A}$ form Lax pair of
Olver-Ibragimov-Shabat type \cite{olver,ibr}. We construct
symplectic and Hamiltonian operators and corresponding Hamiltonian
densities for the two-component \emph{mixed heavenly system} and
\emph{Husain system}. The proof of Jacobi identity is a simple
check that the corresponding symplectic two-forms are closed.
Thus, both systems are set into Hamiltonian form. Applying the
recursion operator $R$ to first Hamiltonian operator, for each
system, we generate explicitly second Hamiltonian operators for
both systems and, thus, we show that they are bi-Hamiltonian
systems, which are integrable in the sense of Magri \cite{magri}.

In section \ref{sec-2comp}, we derive recursion operator for
symmetries and Lax representation for mixed heavenly system.

In section \ref{sec-Hamilton}, we present Lagrangian, symplectic
two-form, symplectic operator and Hamiltonian representation of
the mixed heavenly system.

In section \ref{sec-biHamilton}, using the recursion operator, we
obtain explicitly second and third Hamiltonian representations of
our system and prove that the first two Hamiltonian operators are
compatible, i.e. they form Poisson pencil. Thus, we show that
mixed heavenly equation in a two-component form is an integrable
bi-Hamiltonian system in the sense of Magri.

 In section \ref{sec-symmetries}, we present all point symmetries and
second-order symmetries of the mixed heavenly system and calculate
a table of commutators of symmetry generators. In subsection
\ref{sec-Rsym}, we study the action of the recursion operator on
these symmetries. In subsection \ref{sec-Hamiltonians}, using
inverse Noether theorem for variational symmetries, we determine
conserved densities corresponding to them, which serve as
Hamiltonians of the symmetry flows. In subsection \ref{sec-RH}, we
study recursions of Hamiltonians for symmetry flows of the mixed
heavenly system.

 In section \ref{sec-highflows}, we study hierarchies of mixed
heavenly system and its symmetry flows. In contrast to our
previous studies of bi-Hamiltonian structures of Pleba\~nski
second heavenly equation \cite{nns} and complex Monge-Amp\`ere
equation \cite{nsky}, the mixed heavenly system, though being
bi-Hamiltonian, does not possess an infinite hierarchy but its
hierarchy consists only of two members (the same is true for
Husain system). However, some of its variational symmetries do
form an infinite hierarchy of commuting flows, that contains
higher nonlocal flows. In particular, we present explicitly the
first Hamiltonian flow generating a nonlocal symmetry of our
system. We also obtain bi-Hamiltonian representations for all
local variational symmetry flows.

In section \ref{sec-husain}, we convert Husain equation in a
two-component form and present a Lagrangian, suitable for deriving
Hamiltonian form of the two-component Husain system.

In section \ref{sec-HusHam}, we construct a symplectic two-form,
symplectic and Hamiltonian operators and Hamiltonian density and,
thus, obtain Hamiltonian form of the Husain system.

In section \ref{sec-recurs}, we discover recursion operator and
obtain Lax representation for Husain system.

In section \ref{sec-biHamiltonHus}, we compute second Hamiltonian
operator and prove that the two Hamiltonian operators form Poisson
pencil. Thus, we conclude that Husain equation in a two-component
form is an integrable bi-Hamiltonian system in the sense of Magri
\cite{magri}.

In section \ref{sec-Husymmetries}, we present all point and
second-order symmetries of Husain system and calculate a table of
commutators of symmetry generators. In subsection \ref{sec-RHus},
we study the action of recursion operator on symmetries of Husain
system. In subsection \ref{sec-HamilHus}, we find Hamiltonian
densities of variational symmetry flows which are conserved
densities of the Husain system. In subsection \ref{sec-RH_Hus}, we
study the action of recursion operator on Hamiltonians of the
symmetry flows.

In section \ref{sec-highHus}, we show that the hierarchy of Husain
system consists of two members only. We obtain an infinite
hierarchy of commuting symmetry flows that contains higher
nonlocal flows. We present explicitly the first nonlocal
Hamiltonian flow in this hierarchy. We obtain bi-Hamiltonian
representations for all local variational symmetry flows of Husain
system.

\section{Recursion operator and Lax representation
of mixed heavenly system}
 \setcounter{equation}{0}
 \label{sec-2comp}

By choosing $u_t = v$ as the second unknown, we present mixed
heavenly equation (\ref{mix}) in the form of a two-component
evolution system
\begin{equation}\label{mix2}
 u_t = v,\qquad v_t = \frac{1}{u_{xx}}\big(v_x^2 + v_zu_{xy} -
 v_yu_{xz} + \varepsilon\big)\equiv Q,
\end{equation}
which we shall call \textit{mixed heavenly system}. Lie groups of
symmetry transformations of system (\ref{mix2}) in the canonical
form, when only dependent variables are transformed, are
determined by the Lie equations
\begin{equation}\label{Lie}
  u_\tau = \varphi,\qquad v_\tau = \psi ,
\end{equation}
where $\tau$ is the group parameter. The symmetry condition
amounts to compatibility of Lie equations (\ref{Lie}) and
equations (\ref{mix2}): $u_{t\tau} - u_{\tau t} = 0$ and
$v_{t\tau} - v_{\tau t} = 0$. We introduce a two-component
symmetry characteristic of system (\ref{mix2}): $\Phi = \left(
 \begin{array}{c}
    \varphi
    \\ \psi
 \end{array}
\right)$. Then the symmetry condition results in the linear matrix
equation
\begin{equation}\label{symeq}
  \hat{A}(\Phi) = 0,
\end{equation}
where $\hat{A}$ is the Frech\'et derivative of the flow
(\ref{mix2})
\begin{equation}
 \hat{A} =  \left(\!\!\!\begin{array}{cc} D_t & - 1
\\ \frac{\textstyle Q}{\textstyle u_{xx}}D_x^2
 - \frac{\textstyle v_z}{\textstyle u_{xx}}D_xD_y
+ \frac{\textstyle v_y}{\textstyle u_{xx}}D_xD_z,
 & D_t - \frac{\textstyle 2v_x}{\textstyle u_{xx}}D_x
+ \frac{\textstyle u_{xz}}{\textstyle u_{xx}}D_y
 - \frac{\textstyle u_{xy}}{\textstyle u_{xx}}D_z \end{array}\!\!\!\right)
 \label{A}
\end{equation}
Here $D_t,D_x,D_y,D_z$ are operators of total derivatives with
respect to $t$, $x$, $y$, $z$. In particular, the first row of
(\ref{symeq}) yields $\varphi_t = \psi$. Using this relation and a
similar one for the partner symmetry, $\tilde{\psi} =
\tilde{\varphi_t}$, we rewrite the recursion relation
(\ref{mixrecurs}) with $\omega_0 = 0$ and $u_t = v$, $v_t = Q$ in
the two-component form
\begin{gather}
 \tilde{\psi} = \big(- QD_x + v_{z}D_y -
 v_{y}D_z\big)\varphi + v_{x}\psi ,\notag
\\ \tilde{\varphi}_x = \big(- v_{x}D_x + u_{xz}D_y -
 u_{xy}D_z\big)\varphi + u_{xx}\psi .
 \label{sym2}
\end{gather}
After integrating the second equation (\ref{sym2}) with respect to
$x$ at constant $y,z$ and $t$ and using the notation $\tilde{\Phi}
= \left(
 \begin{array}{c}
    \tilde{\varphi}
    \\ \tilde{\psi}
 \end{array}
\right)$, the recursion relation takes the matrix form
$\tilde{\Phi} = R\big(\Phi\big)$, where the recursion operator $R$
is the $2\times 2$ matrix
\begin{equation}\label{R}
 R = \left(
  \begin{array}{lc}
 D_x^{-1}\Big(- v_xD_x + u_{xz}D_y - u_{xy}D_z\Big)
 &  D_x^{-1}u_{xx}
 \\[2mm] - QD_x + v_zD_y - v_yD_z & v_x
  \end{array}
  \right).
\end{equation}

For the commutator of the recursion operator $R$ and operator
$\hat{A}$ of the symmetry condition (\ref{symeq}), computed
without using the equations of motion, we obtain
\begin{gather}\label{commut}
 [R , \hat{A}] =
\\[3mm] \left[
\begin{array}{ll}
 \begin{array}{l}
  (v_t - Q)_x - D_x^{-1}\big[(v_t - Q)_{xx}
  \\  + (u_t - v)_{xz}D_y - (u_t - v)_{xy}D_z\big]
  \end{array}
  & - D_x^{-1}(u_t - v)_{xx}
 \\[10mm]  \begin{array}{l}
 \frac{\textstyle 1}{\textstyle u_{xx}}\bigl[
 - Q(u_t - v)_{xx} + v_z(u_t - v)_{xy}
 \\  - v_y(u_t - v)_{xz} + 2v_x(v_t - Q)_x - u_{xz}(v_t - Q)_y
 \\  + u_{xy}(v_t - Q)_z \bigr]D_x  - (v_t - Q)_zD_y + (v_t - Q)D_z
      \end{array}
& - (v_t - Q)_x
\end{array}
 \right] \notag
\end{gather}
Therefore, on solutions of the system (\ref{mix2}) operators $R$
and $\hat{A}$ commute and hence $R$ acting on any two-component
symmetry generates again a symmetry. This proves that $R$ is
indeed a recursion operator. Moreover, vanishing of the commutator
(\ref{commut}) reproduces (\ref{mix2}) and hence operators $R$ and
$\hat{A}$ form a Lax pair of the Olver-Ibragimov-Shabat type
\cite{olver,ibr} for mixed heavenly system (\ref{mix2}).

\section{Lagrangian, symplectic and Hamiltonian\\ structure
         of mixed heavenly system}
\setcounter{equation}{0}
 \label{sec-Hamilton}

We start with the Lagrangian for the mixed heavenly system
(\ref{mix2})
\begin{equation}\label{lagrange}
  L = \left(vu_t - \frac{1}{2}\,v^2\right)u_{xx} +
  \frac{1}{3}\,u_t(u_yu_{xz} - u_zu_{xy}) + \varepsilon u ,
\end{equation}
which yields the canonical momenta
\begin{equation}\label{moment}
\pi_u = \frac{\partial L}{\partial u_t} = vu_{xx} +
 \frac{1}{3}\,(u_yu_{xz} - u_zu_{xy}),\quad
\pi_v = \frac{\partial L}{\partial v_t} = 0 ,
\end{equation}
that cannot be inverted for the velocities $u_t$ and $v_t$, and
therefore the Lagrangian (\ref{lagrange}) is degenerate. Following
Dirac's theory of constraints \cite{dirac}, we treat the
definitions (\ref{moment}) as constraints of the second class
\begin{equation}\label{constr}
  \phi_u = \pi_u - vu_{xx} - \frac{1}{3}\,(u_yu_{xz} - u_zu_{xy})
  = 0, \quad \phi_v = \pi_v = 0 ,
\end{equation}
compute the Poisson brackets of the constraints (for details of
this procedure see \cite{nns})
\begin{equation}\label{poiss}
[\phi_i(x,y,z),\phi_j(x',y',z')] = K_{ij},\qquad i,j = 1,2, \qquad
u_1 =u ,\; u_2 = v
\end{equation}
as entries of the $2\times 2$ matrix
\begin{equation}\label{K}
 K =  \left( \begin{array}{cc}
 D_xv_x + v_xD_x + \frac{1}{2}\left(D_zu_{xy} + u_{xy}D_z\right)
 - \frac{1}{2}\left(D_yu_{xz} + u_{xz}D_y\right) & - u_{xx} \\
  u_{xx} & 0
\end{array}  \right)
\end{equation}
which is an explicitly skew-symmetric symplectic operator. The
corresponding symplectic two-form is a volume integral $\Omega =
\int\limits_{V}\omega dx dy dz$ of the density
\begin{gather}
 \omega = \frac{1}{2} \, d u^i \wedge K_{ij} \, d u^j
  \label{sympform}
 \\ \qquad = v_x d u\wedge d u_x - u_{xx}d u\wedge d v
 + \frac{1}{2} \left(u_{xy} du \wedge du_z - u_{xz} d u\wedge d u_y
  \right). \notag
\end{gather}
The form $\Omega$ is closed since the exterior differential of
(\ref{sympform}) is a total divergence
\begin{gather}
  d\omega = \frac{1}{2}\,d u_x\wedge d u_y \wedge d u_z \notag
\\ \quad \mbox{} = \frac{1}{6}\Bigl( D_x(du \wedge du_y \wedge du_z) +
D_y(du_x \wedge du \wedge du_z) + D_z(du_x \wedge du_y \wedge
du)\Bigr),
 \label{domeg}
\end{gather}
which is equivalent to zero under the volume integral in $\Omega$,
at appropriate boundary conditions. Therefore, $\Omega$ is indeed
a symplectic form and so $K$, defined by (\ref{K}), is indeed a
symplectic operator. Hence its inverse is a Hamiltonian operator
\begin{equation}\label{J0}
 J_0 = K^{-1} = \left(
\begin{array}{cc}
0 & \frac{\textstyle 1}{\textstyle u_{xx}}
\\[4mm] -\frac{\textstyle 1}{\textstyle u_{xx}} &
   \begin{array}{c}
\frac{\textstyle v_x}{\textstyle u_{xx}} D_x\frac{\textstyle
1}{\textstyle u_{xx}} + \frac{\textstyle 1}{\textstyle
u_{xx}}D_x\frac{\textstyle v_x}{\textstyle u_{xx}} -
\frac{\textstyle u_{xz}}{\textstyle 2u_{xx}} D_{y}\frac{\textstyle
1}{\textstyle u_{xx}}
 \\[3mm] - \frac{\textstyle 1}{\textstyle 2u_{xx}}D_y\frac{\textstyle u_{xz}}{\textstyle
u_{xx}} + \frac{\textstyle u_{xy}}{\textstyle 2u_{xx}}
D_{z}\frac{\textstyle 1}{\textstyle u_{xx}} + \frac{\textstyle
1}{\textstyle 2u_{xx}}D_z\frac{\textstyle u_{xy}}{\textstyle
u_{xx}}
    \end{array}
\end{array}
\right),
\end{equation}
since it is explicitly skew-symmetric and Jacobi identity is
satisfied as a consequence of the closeness of symplectic two-form
$\Omega$.

The Hamiltonian density, corresponding to $J_0$, is defined as
\[H_1 = \pi_u u_t + \pi_v v_t - L\]
 with the result
\begin{equation}
 H_1 = \frac{1}{2}\,v^2u_{xx} - \varepsilon u \quad \iff \quad
 H_1 = \frac{1}{2}\,(v^2 - \varepsilon x^2)u_{xx} ,
 \label{H1}
\end{equation}
where equivalent Hamiltonian densities differ only by a total
$x$-derivative, which vanishes in the Hamiltonian ${\cal
H}^i=\iiint_{-\infty}^{+\infty} H^i dx dy dz$ due to appropriate
boundary conditions at infinity. Indeed,
 \[\frac{1}{2}\,x^2u_{xx} = u + D_x\left(\frac{1}{2}\,x^2u_x - xu\right)
  \iff u.\]

Thus, the mixed heavenly equation in two-component form
(\ref{mix2}) can be presented as the Hamiltonian system
\begin{equation}
\left(
\begin{array}{c}
u_t \\ v_t
\end{array}
\right) =  J_0 \left(
\begin{array}{c}
\delta_u H_1 \\ \delta_v H_1
\end{array}
\right),
 \label{Hamilton}
\end{equation}
where $\delta_u$ and $\delta_v$ are Euler-Lagrange operators
\cite{olv} with respect to $u$ and $v$ applied to the Hamiltonian
density $H_1$ (they correspond to variational derivatives of the
Hamiltonian functional $\int\limits_V H_1 dV$).

\section{Bi-Hamiltonian representation of mixed\\ heavenly system}
\setcounter{equation}{0}
 \label{sec-biHamilton}

By a theorem of Magri \cite{magri}, we can generate second
Hamiltonian operator by acting with the recursion operator
(\ref{R}) on the Hamiltonian operator (\ref{J0})
\begin{equation}\label{J1}
  J_1 = RJ_0 = \left(
  \begin{array}{cc}
   - D_x^{-1} & \frac{\textstyle v_x}{\textstyle u_{xx}}
   \\[3mm] - \frac{\textstyle v_x}{\textstyle u_{xx}} & J_1^{22}
  \end{array}
  \right),
\end{equation}
where $J_1^{22}$, in an explicitly skew-symmetric form, is defined
as
\begin{gather}
  J_1^{22} = \frac{1}{2}\left\{- \left(Q_-D_x\frac{\textstyle 1}{\textstyle u_{xx}}
 + \frac{\textstyle 1}{\textstyle u_{xx}}D_xQ_-\right) +
 \left(v_zD_y\frac{\textstyle 1}{\textstyle u_{xx}} + \frac{\textstyle 1}{\textstyle
 u_{xx}}D_yv_z \right)\right. \notag
 \\[2mm] \qquad \left. - \left(v_yD_z\frac{\textstyle 1}{\textstyle u_{xx}} + \frac{\textstyle 1}{\textstyle
 u_{xx}}D_zv_y \right) - \left(\frac{\textstyle v_x}{\textstyle u_{xx}}D_y\frac{\textstyle u_{xz}}{\textstyle u_{xx}}
 + \frac{\textstyle u_{xz}}{\textstyle u_{xx}}D_y \frac{\textstyle v_x}{\textstyle
 u_{xx}}\right) \right.\notag
 \\[2mm] \qquad \qquad \qquad \qquad \qquad \qquad \qquad
 \left. + \left(\frac{\textstyle v_x}{\textstyle u_{xx}}D_z\frac{\textstyle u_{xy}}{\textstyle u_{xx}}
 + \frac{\textstyle u_{xy}}{\textstyle u_{xx}}D_z \frac{\textstyle v_x}{\textstyle
 u_{xx}}\right)
 \right\},
 \label{J1^22}
\end{gather}
where we have denoted
 \[Q_- = Q - \frac{2v_x^2}{u_{xx}} = \frac{1}{u_{xx}}\big(- v_x^2 + v_zu_{xy} -
 v_yu_{xz} + \varepsilon\big) .\]
 The proof of Jacobi identity
is straightforward and lengthy. The calculations can be simplified
by using P. Olver's criterion (theorem 7.8 in \cite{olv})
formulated in terms of functional multivectors.

We have also performed a straightforward check for compatibility
of two Hamiltonian operators $J_0$ and $J_1$ using P. Olver's
criterion (corollary 7.21 in his book \cite{olv}) and proved that
every linear combination $aJ_0 + bJ_1$ with arbitrary constant
coefficients $a$ and $b$ satisfies Jacobi identity, i.e. $J_0$ and
$J_1$ form Poisson pencil (called Hamiltonian pair in \cite{olv}).

The mixed heavenly flow (\ref{mix2}) can be generated by
Hamiltonian operator $J_1$ from the Hamiltonian density
\begin{equation}\label{H0}
  H_0 = (c - x)vu_{xx} ,
\end{equation}
where $c$ is a constant, so that the mixed heavenly equation in
the two-component form (\ref{mix2}) admits two Hamiltonian
representations
\begin{equation}\label{biHam}
\left(
\begin{array}{c}
u_t \\ v_t
\end{array}
\right) =  J_0 \left(
\begin{array}{c}
\delta_u H_1 \\ \delta_v H_1
\end{array}
\right) =  J_1 \left(
\begin{array}{c}
\delta_u H_0 \\ \delta_v H_0
\end{array}
\right)
\end{equation}
and thus it is a {\it bi-Hamiltonian system}.

We note that we could drop out the term $h_0 = cvu_{xx}$ in the
Hamiltonian $H_0$ and set $c=0$ in (\ref{H0}), so that $H_0 = x v
u_{xx}$, because the vector of variational derivatives of $h_0$
belongs to the kernel of $J_1$
\begin{equation}\label{kerJ1}
 J_1 \left(
\begin{array}{c}
\delta_u h_0 \\ \delta_v h_0
\end{array}
\right) = \left(
\begin{array}{c}
0 \\ 0
\end{array}
\right).
\end{equation}

According to Magri's theorem, by repeated applications of
recursion operator to the first Hamiltonian operator $J_0$, we
could generate an infinite sequence of Hamiltonian operators
\begin{equation}\label{Jn}
  J_n = R^n J_0 .
\end{equation}
In particular, for $n=2$ we obtain a new Hamiltonian operator $J_2
= R^2 J_0 = R J_1$, which has the following explicitly
skew-symmetric form
\begin{equation}
J_{2} =   \left(  \begin{array}{cc}
           \begin{array}{c}
\frac{1}{2} D_{x}^{-1}(u_{xy}D_z + D_zu_{xy} \\ - u_{xz}D_y -
D_yu_{xz})D_x^{-1}
           \end{array}
 &\hspace*{7mm}
           \begin{array}{c}
 D_x^{-1}(D_yv_z - D_zv_y) \\ + \frac{\textstyle 1}{\textstyle u_{xx}
 }\,(v_yu_{xz} - v_zu_{xy} - \varepsilon)
           \end{array}
 \\[8mm]
           \begin{array}{c}
- \left\{(v_zD_y - v_yD_z)D_x^{-1} \right. \\ \left. +
\frac{\textstyle 1}{\textstyle u_{xx}
 }\,(v_yu_{xz} - v_zu_{xy} - \varepsilon)\right\}
           \end{array} &\hspace*{7mm} J_2^{22}
 \end{array}
  \right)
  \label{J2}
\end{equation}
where $J_2^{22}$ is defined by
\begin{gather}
  J_2^{22} = \frac{v_x}{u_{xx}^2}\left\{(v_yu_{xz} - v_zu_{xy} -
 \varepsilon)D_x + \left(v_zu_{xx} - \frac{1}{2}v_xu_{xz}\right)D_y
 \right. \notag
\\ \qquad\qquad \left. - \left(v_yu_{xx} - \frac{1}{2}v_xu_{xy}\right)D_z
\right\}  +  \Bigg\{D_x(v_yu_{xz} - v_zu_{xy} - \varepsilon)
\notag
\\ \qquad\qquad\qquad\qquad \mbox{} + D_y\left(v_zu_{xx} - \frac{1}{2}v_xu_{xz}\right) - D_z
\left(v_yu_{xx} - \frac{1}{2}v_xu_{xy}\right)
\Bigg\}\frac{v_x}{u_{xx}^2}.
 \label{J2^22}
\end{gather}
Surprisingly enough, the Hamiltonian density, from which the
Hamiltonian operator $J_2$ generates the system (\ref{mix2}), is
proportional to $H_1$ defined by (\ref{H1})
\begin{equation}\label{H_{-1}}
H_{-1} = -\varepsilon H_1 = u - \frac{\varepsilon}{2}\,v^2 u_{xx},
\end{equation}
so that the mixed heavenly system admits the three-Hamiltonian
representation
\begin{equation}\label{Ham3}
\left(
\begin{array}{c}
u_t \\ v_t
\end{array}
\right) =  J_0 \left(
\begin{array}{c}
\delta_u H_1 \\ \delta_v H_1
\end{array}
\right) =  J_1 \left(
\begin{array}{c}
\delta_u H_0 \\ \delta_v H_0
\end{array}
\right) = J_2 \left(
\begin{array}{c}
\delta_u (H_{-1}) \\ \delta_v (H_{-1})
\end{array}
\right).
\end{equation}
The result (\ref{H_{-1}}) for $H_{-1}$ is derived in a regular way
in (\ref{H0tild}) (at $c(y,z) = 0$) in subsection \ref{sec-RH}
(see also (\ref{R+})).

Computing $J_n$ in (\ref{Jn}) for $n = 3,4,\ldots$, we can
generate an infinite series of Hamiltonian operators, which shows
that the mixed heavenly equation, considered in a two-component
form, is a multi-Hamiltonian system in the above-mentioned sense.

\section{Symmetries and conservation laws for\\ mixed heavenly system}
\setcounter{equation}{0}
 \label{sec-symmetries}

Using the software packages LIEPDE and CRACK by T. Wolf
\cite{wolf}, run under REDUCE 3.8, we have calculated all point
and Lie-B\"acklund second-order symmetries of mixed heavenly
system (\ref{mix2}). For point symmetries, we list generators and
two-component symmetry characteristics $\Phi = (\varphi, \psi)^T$,
where $T$ means transpose:
\begin{gather}
 X_1 = t\partial_u + \partial_v, \quad \varphi_1 = t,\quad
\psi_1 = 1
 \notag
\\ X_2 = - \partial_x, \quad \varphi_2 = u_x ,\quad \psi_2 = v_x
 \notag
  \\ X_3^{a} = a(y,z)\partial_u, \quad \varphi_{3a} = a(y,z),\quad
  \psi_{3a} = 0
  \label{point}
    \\ X_4^{a} = a_y(y,z)\partial_z - a_z(y,z)\partial_y,
    \quad \varphi_{4a} = a_zu_y - a_yu_z,\quad
  \psi_{4a} = a_zv_y - a_yv_z
  \notag
  \\ X_5 = - \left\{\frac{1}{2}(y\partial_y + z\partial_z) + t\partial_t +
  u\partial_u\right\},
  \notag
  \\ \varphi_5 = tv - u + \frac{1}{2}(yu_y + zu_z),\qquad
  \psi_5 = tQ + \frac{1}{2}(yv_y + zv_z) \notag
  \\ X_6 = t\partial_t - x\partial_x - v\partial_v,\quad
  \varphi_6 = xu_x - tv,\quad \psi_6 = xv_x - tQ - v
  \notag
  \\ X_{c(x,v)} = - c_v(x,v)\partial_t + (c - vc_v)\partial_u,
  \quad \varphi_c = c(x,v),\quad \psi_c = c_v(x,v)Q,
  \notag
\end{gather}
where $c(x,v)$ is an arbitrary smooth solution of the equation
\begin{equation}\label{c_eq}
  c_{xx}(x,v) + \varepsilon c_{vv}(x,v) = 0
\end{equation}
and we have used the equations of motion (\ref{mix2}) for
eliminating $u_t$ and $v_t$. Such obvious symmetries as
translations in $y,z$ and $u$, which do not appear explicitly in
the list (\ref{point}), can be obtained as simple particular cases
of $X_3^a$ and $X_4^a$, while translations in $t$ are obtained
from $X_c$ at $c = - v$.

All second-order Lie-B\"acklund symmetries \cite{ibragim} modulo
point symmetries have generators of the form
\begin{gather}
  \hat{X}_a = a(t,x,v,u_x)\partial_u
  + (a_t + a_vQ + a_{u_x}v_x)\partial_v + \cdots, \notag
  \\ \varphi_a = a(t,x,v,u_x),\qquad \psi_a = a_t + a_vQ +
  a_{u_x}v_x,
  \label{order2}
\end{gather}
where the dots denote an infinite prolongation part and
$a(t,x,v,u_x)$ is an arbitrary smooth solution of the equations
\begin{gather}
  a_{tx} - \varepsilon a_{vu_x} = 0,\quad a_{tv} + a_{xu_x} = 0
,\quad a_{xx} + \varepsilon a_{vv} = 0,\quad a_{tt} + \varepsilon
a_{u_xu_x} = 0. \notag \\
  \label{aeq}
\end{gather}
Corresponding Lie equations have the form of a second-order flow,
due to the definition of $Q$ in (\ref{mix2}),
\begin{equation}\label{Lieord2}
  u_\tau = a(t,x,v,u_x),\quad v_\tau = a_t + a_vQ +
  a_{u_x}v_x
\end{equation}
where the ``time'' $\tau$ is the group parameter. We note that
mixed heavenly system itself in (\ref{mix2}) is a particular case
of (\ref{Lieord2}) at $a=v$, that obviously satisfies conditions
(\ref{aeq}). We also note that $(\varphi_6, \psi_6)$ correspond to
the particular case of second-order symmetries (\ref{order2}) with
$a = xu_x - tv$. We point out that the symmetry characteristic
$(\varphi_c, \psi_c)$ in (\ref{point}) together with the condition
(\ref{c_eq}) is a particular case of the symmetry $(\varphi_a,
\psi_a)$ in (\ref{order2}) with $a = c(x,v)$ satisfying the
conditions (\ref{aeq}). Similarly, we see that symmetry generators
$X_1$ and $X_2$ in evolutionary form are also particular cases of
the second-order symmetry generator $\hat{X}_{a(t,x,v,u_x)}$ while
$X_3^{a(y,z)}$, $X_4^{a(y,z)}$ and $X_5$ are not.

All second-order Lie-B\"acklund symmetries can be obtained by
taking linear combinations of the generators (\ref{order2}) and
point symmetries generators $X_3^{a(y,z)}$, $X_4^{a(y,z)}$ and
$X_5$.

\begin{table}[ht]
\caption{Commutators of symmetries of mixed heavenly system.}
\rule[2mm]{13.4cm}{1pt} \\
{\begin{tabular}{lcc@{\hspace{-.05pt}}ccc@{\hspace{-.05pt}}c@{\hspace{2pt}}c@{\hspace{1pt}}c}
              &$X_1$&$X_2$&$X_3^{f(y,z)}$ & $X_4^{b(y,z)}$& $X_5$
              & $X_6$ & $X_{c(x,v)}$ & $\hat{X}_{b(t,x,v,u_x)}$
              \\[2pt] \hline \\[-2pt]
 $X_1$        &  0 & 0  & 0              & 0             & 0&$-X_1$& $X_{c_v}$ & $\hat{X}_{b_v}$ \\
 $X_2$        &  0 & 0  & 0             & 0              & 0&$-X_2$& $-X_{c_x}$& $-\hat{X}_{b_x}$  \\
 $X_3^{a(y,z)}$& 0 & 0  & 0             & $X_3^{\frac{\partial(a,b)}{\partial(y,z)}}$& $\frac{1}{2}X_3^{\hat{a}(y,z)}$
 & 0 & 0 & 0 \\
 $X_4^{a(y,z)}$& 0 & 0  &$-X_3^{\frac{\partial(f,a)}{\partial(y,z)}}$ & $X_4^{\frac{\partial(a,b)}{\partial(y,z)}}$ &
 $\frac{1}{2}X_4^{\hat{a}(y,z)}$& 0 & 0 & 0 \\
 $X_5$         & 0 & 0  & $-\frac{1}{2}X_3^{\hat{f}}$ & $-\frac{1}{2}X_4^{\hat{b}}$& 0 & 0
 & $X_c$ & $-\hat{X}_{b^{\,\prime}}$ \\
 $X_6$   & $X_1$ & $X_2$ & 0 & 0 & 0 & 0 & $X_{\tilde{c}}$ & $\hat{X}_{\tilde{b}}$ \\
 $X_{d(x,v)}$ & $-X_{d_v}$ & $X_{d_x}$ & 0 & 0 & $-X_{d}$ & $-X_{\tilde{d}}$
              & 0 & $-\hat{X}_{\langle b,d\rangle}$ \\
 $\hat{X}_{a(t,x,v,u_x)}$ & $-\hat{X}_{a_v}$ & $\hat{X}_{a_x}$ & 0
 & 0 & $\hat{X}_{a^{\,\prime}}$ & $-\hat{X}_{\tilde{a}}$ &
 $\hat{X}_{\langle a,c\rangle}$ & $\hat{X}_{\ll a,b\gg}$
\end{tabular}}
\\ \rule[.5pt]{13.4cm}{1pt}
\end{table}

In table \thetable\ we present commutators of symmetry generators,
where the commutator $\left[X_i,X_j\right]$ is given at the
intersection of $i$th row and $j$th column. We have used here the
following shorthand notation:
\begin{gather}
 \frac{\partial(a,b)}{\partial(y,z)} = a_yb_z - a_zb_y, \qquad \langle a,c\rangle = a_tc_v - a_{u_x}c_x, \notag
 \\ \ll a,b\gg = \frac{\partial(a,b)}{\partial(t,v)} +
 \frac{\partial(a,b)}{\partial(x,u_x)}, \quad \tilde{a} = ta_t -
 xa_x + u_xa_{u_x} - va_v,
 \notag
 \\ \hat{a} = ya_y + za_z - 2a,\quad a^{\,\prime} = ta_t + u_xa_{u_x} - a.
 \label{not}
\end{gather}
Let $\hat{X}_\Phi$ be an evolution form of a symmetry generator
with the characteristic $\Phi = \left(\begin{array}{c}
 \varphi \\ \psi
\end{array}\right)$. Consider the evolution system
of PDEs
\begin{equation}\label{evsys}
 \left(\begin{array}{c}
 u_t \\ v_t
\end{array}\right) = F([u],[v],\vec{r}) \equiv \left(\begin{array}{c}
 f \\ g
\end{array}\right),
\end{equation}
where $\vec{r} = (x,y,z)$ and $[u]$, $[v]$ denote unknown
functions $u$ and $v$ of $\vec{r}$ together with their partial
derivatives with respect to the components of $\vec{r}$. Let
$\hat{X}_F$ be an evolutionary generator of the flow (\ref{evsys})
with the characteristic $F = \left(\begin{array}{c}
 f \\ g
\end{array}\right)$. Then \textit{$\hat{X}_\Phi$ generates a symmetry of the
flow (\ref{evsys})} if and only if the following commutator
relation is satisfied \cite{ff,sheftel}
\begin{equation}\label{symmcomm}
  \left[\hat{X}_\Phi, \hat{X}_F\right ] = \hat{X}_{\Phi_t},
\end{equation}
where $\Phi_t$ denotes partial derivative of $\Phi$ with respect
to $t$. In particular, if $\Phi$ does not depend explicitly on
time $t$, then these two generators must commute:
$\left[\hat{X}_\Phi, \hat{X}_F\right] = 0$. For mixed heavenly
system (\ref{mix2}) we have $F = (v,Q)^T$ and then the results of
table \thetable\ for $\hat{X}_b = \hat{X}_v$ show that equation
(\ref{symmcomm}) is indeed satisfied for all the generators in the
left column of the table, which provides an independent test for
all of the symmetries of (\ref{mix2}).

\subsection{Action of recursion operator on symmetries of the
mixed heavenly system}
 \label{sec-Rsym}

In order to obtain correctly the action of recursion operator $R$
on the space of symmetries, we note that the operator $D_x^{-1}$
in $R$ should be understood as an indefinite integral with respect
to $x$ with the ``constant'' of integration $C(y,z,t)$. Indeed,
the integrability condition for the recursion relation
(\ref{mixrecurs}) in a two-component notation (\ref{sym2}) has the
form $\tilde{\varphi}_{xt} = \tilde{\psi}_x$, which implies
$\tilde{\varphi}_{t} = \tilde{\psi} + C(y,z,t)$ with an arbitrary
function $C(y,z,t)$. Therefore, the required relation
$\tilde{\varphi}_{t} = \tilde{\psi}$ does not follow from
(\ref{sym2}) but may present an additional constraint on the
$t$-dependence of $C(y,z,t)$ when $R$ is applied to symmetries
$\Phi = \left(
 \begin{array}{c}
    \varphi
    \\ \psi
 \end{array}
\right)$, while the additive term $c(y,z)$ in $C(y,z,t)$ will
still be completely arbitrary. In the Lax representation
(\ref{commut}) and definition (\ref{J1}) of $J_1$ we have to
choose $c(y,z) = 0$.

 Proceeding to the action of the recursion operator on symmetries,
we start with the recursion
\begin{equation}\label{Rfi1}
\left( \begin{array}{c}
    \tilde{\varphi}_1
    \\ \tilde{\psi}_1
 \end{array}
\right)
=
  R\left(
 \begin{array}{c}
    \varphi_1
    \\ \psi_1
 \end{array}
\right) = R\left(
 \begin{array}{c}
    t
    \\ 1
 \end{array}
\right) = \left(
 \begin{array}{c}
    D_x^{-1}u_{xx}
    \\ v_x
 \end{array}
\right) = \left(
 \begin{array}{c}
    u_x + C(y,z,t)
    \\ v_x
 \end{array}
\right),
\end{equation}
but due to the constraint $\tilde{\varphi}_{t} = \tilde{\psi}$ we
have $C_t = 0$ and so $C = c(y,z)$ with an arbitrary function
$c(y,z)$. Finally we have
\begin{equation}\label{Rfi1fin}
\left( \begin{array}{c}
    \tilde{\varphi}_1
    \\ \tilde{\psi}_1
 \end{array}
\right) =
 \left(
 \begin{array}{c}
    u_x + c(y,z)
    \\ v_x
 \end{array}
\right) = \left(
 \begin{array}{c}
    \varphi_2
    \\ \psi_2
 \end{array}
\right) + \left(
 \begin{array}{c}
    \varphi_{3c}
    \\ \psi_{3c}
 \end{array}
\right),
\end{equation}
where $\left(
 \begin{array}{c}
    \varphi_{3c}
    \\ \psi_{3c}
 \end{array}
\right) = \left(
 \begin{array}{c}
    c(y,z)
    \\ 0
 \end{array}
\right)$. Our next step is to compute
\begin{equation}\label{Rfi2}
\left( \begin{array}{c}
    \tilde{\varphi}_2
    \\ \tilde{\psi}_2
 \end{array}
\right) =
  R\left(
 \begin{array}{c}
    \varphi_2
    \\ \psi_2
 \end{array}
\right) = R\left(
 \begin{array}{c}
    u_x
    \\ v_x
 \end{array}
\right) = \left(
 \begin{array}{c}
    D_x^{-1}(0)
    \\ - \varepsilon
 \end{array}
\right) = \left(
 \begin{array}{c}
    C(y,z,t)
    \\ - \varepsilon
 \end{array}
\right)
\end{equation}
using in $R$ the definition of $Q$ from (\ref{mix2}). Here the
``constant'' of integration $C(y,z,t)$ could not be neglected
since the choice $C=0$ will violate our constraint
$\tilde{\varphi}_{t} = \tilde{\psi}$, so that $\left(
 \begin{array}{c}
    0
    \\ - \varepsilon
 \end{array}
\right)$ will not be a symmetry. Instead, the constraint yields
$C_t(y,z,t) = - \varepsilon$ so that $C = - \varepsilon t +
c(y,z)$ with an arbitrary function $c(y,z)$, which implies the
result
\begin{equation}\label{Rfi2fin}
\left( \begin{array}{c}
    \tilde{\varphi}_2
    \\ \tilde{\psi}_2
 \end{array}
\right) =
 - \varepsilon \left(
 \begin{array}{c}
    t
    \\ 1
 \end{array}
\right) + \left(
 \begin{array}{c}
    c(y,z)
    \\ 0
 \end{array}
\right) = - \varepsilon \left(
 \begin{array}{c}
    \varphi_1
    \\ \psi_1
 \end{array}
\right) + \left(
 \begin{array}{c}
    \varphi_{3c}
    \\ \psi_{3c}
 \end{array}
\right).
\end{equation}
Similarly, we obtain
\begin{equation}\label{Rfi3}
\left( \begin{array}{c}
    \tilde{\varphi}_{3a}
    \\ \tilde{\psi}_{3a}
 \end{array}
\right) =
  R\left(
 \begin{array}{c}
    \varphi_{3a}
    \\ \psi_{3a}
 \end{array}
\right) = R\left(
 \begin{array}{c}
    a(y,z)
    \\ 0
 \end{array}
\right) = \left(
 \begin{array}{c}
    \varphi_{3c}
    \\ \psi_{3c}
 \end{array}
\right) - \left(
 \begin{array}{c}
    \varphi_{4a}
    \\ \psi_{4a}
 \end{array}
\right).
\end{equation}
  At the next step we obtain the first nonlocal symmetry
\begin{gather}
\left( \begin{array}{c}
    \tilde{\varphi}_{4a}
    \\ \tilde{\psi}_{4a}
 \end{array}
\right) =
  R\left(
 \begin{array}{c}
    \varphi_{4a}
    \\ \psi_{4a}
 \end{array}
\right)
 =
  R\left(
 \begin{array}{c}
    a_zu_y - a_yu_z
    \\ a_zv_y - a_yv_z
 \end{array}
\right) \notag
 \\
 = \left(
\begin{array}{c}
  D_x^{-1}\left\{u_{xx}\psi_{4a} + u_{xz}(a_yv_x + \varphi_{4a,y}) -
  u_{xy}(a_zv_x + \varphi_{4a,z})\right\}
  \\[1mm] v_z(\varphi_{4a,y} - a_yv_x) - v_y(\varphi_{4a,z} - a_zv_x) -
  Q\varphi_{4a,x}
 \end{array}
\right)
 \label{Rfi4}
\end{gather}
with $\varphi_{4a,y} = D_y(\varphi_{4a})$ and so on, since the
expression in curly braces in the first row is not a total
$x$-derivative. Another nonlocal symmetry is obtained by the
action of $R$ on $X_5$:
\begin{eqnarray}
& & \left( \begin{array}{c}
    \tilde{\varphi}_5
    \\ \tilde{\psi}_5
 \end{array}
\right) =
  R\left(
 \begin{array}{c}
    \varphi_5
    \\ \psi_5
 \end{array}
\right)
 \label{Rfi5}
 \\ & & \qquad\qquad = \frac{1}{2} \left( \begin{array}{c}
 D_x^{-1}\{u_{xx}(yv_y + zv_z) - v_xw_x + v_zw_y - v_yw_z\} +
 \varepsilon tx
 \\ v_x(yv_y + zv_z) - Qw_x + v_zw_y - v_yw_z
 \end{array}
\right), \nonumber
\end{eqnarray}
where we have denoted $w = yu_y + zu_z - 2u$. The action of $R$ on
$X_6$ yields
\begin{equation}\label{Rfi6}
\left( \begin{array}{c}
    \tilde{\varphi}_6
    \\ \tilde{\psi}_6
 \end{array}
\right) =
  R\left(
 \begin{array}{c}
    \varphi_6
    \\ \psi_6
 \end{array}
\right) =
 - \left(
 \begin{array}{c}
    u_xv + \varepsilon xt
    \\ u_xQ + vv_x + \varepsilon x
 \end{array}
\right) + \left(
 \begin{array}{c}
    c(y,z)
    \\ 0
 \end{array}
\right),
\end{equation}
where we have again used the constraint $\tilde{\varphi}_{6\,t} =
\tilde{\psi}_{6}$ on the ``constant'' of integration $C(y,z,t)$.
Finally we consider the action of $R$ on the symmetry $X_c$ in
(\ref{point}) with $c(x,v)$ satisfying (\ref{c_eq}):
\begin{equation}\label{Rfic}
\left( \begin{array}{c}
    \tilde{\varphi}_c
    \\ \tilde{\psi}_c
 \end{array}
\right) =
  R\left(
 \begin{array}{c}
    \varphi_c
    \\ \psi_c
 \end{array}
\right) =
 R\left(
 \begin{array}{c}
    c(x,v)
    \\ c_v(x,v)Q
 \end{array}
\right) =
 \left(
 \begin{array}{c}
    D_x^{-1}(-v_xc_x + \varepsilon c_v)
    \\ - c_x Q
 \end{array}
\right).
\end{equation}
Equation (\ref{c_eq}) in the form $(c_x)_x = - \varepsilon
(c_v)_v$ implies local existence of the function $f(x,v)$ such
that $f_v = c_x$ and $f_x = - \varepsilon c_v$ and hence $f(x,v)$
satisfies the same equation as $c(x,v)$: $f_{xx} + \varepsilon
f_{vv} = 0.$ The result (\ref{Rfic}), being expressed in terms of
$f(x,v)$, becomes
\begin{equation}\label{Rficf}
\left( \begin{array}{c}
    \tilde{\varphi}_c
    \\ \tilde{\psi}_c
 \end{array}
\right) = - \left(
 \begin{array}{c}
    D_x^{-1}D_x[f(x,v)]
    \\  f_v Q
 \end{array}
\right) = - \left(
 \begin{array}{c}
        f(x,v)
    \\  f(x,v)Q
 \end{array}
\right) + \left(
 \begin{array}{c}
  a(y,z) \\ 0
\end{array}
\right),
\end{equation}
where $a(y,z)$ is the ``constant'' of integration. If we neglect
$a(y,z)$, then the result (\ref{Rficf}) reads $\left(
 \begin{array}{c}
    \tilde{\varphi}_c
    \\ \tilde{\psi}_c
 \end{array}
\right) = - \left(
 \begin{array}{c}
  \varphi_f
  \\ \psi_f
 \end{array}
\right)$, so that recursion acts on the solution space of equation
(\ref{c_eq}).

Next we consider action of $R$ on second-order symmetries
$\hat{X}_a$ in (\ref{order2}) with $a(t,x,v,u_x)$ that satisfies
four linear equations (\ref{aeq}):
\begin{eqnarray}
 \hspace*{-4cm} & &  \left( \begin{array}{c}
    \tilde{\varphi}_a
    \\ \tilde{\psi}_a
 \end{array}
\right) =
  R\left(
 \begin{array}{c}
    \varphi_a
    \\ \psi_a
 \end{array}
\right) =
  R\left(
 \begin{array}{c}
  a(t,x,v,u_x)
    \\  a_t + a_vQ + a_{u_x}v_x
 \end{array}
\right) \nonumber
 \\ \hspace*{-4cm} & & \qquad\qquad\qquad\qquad\quad\; = \left(
 \begin{array}{c}
  D_x^{-1}(a_tu_{xx} - a_xv_x + \varepsilon a_v)
    \\ - \varepsilon a_{u_x} - a_x Q + a_tv_x
 \end{array}
\right).
 \label{Rfiord2}
\end{eqnarray}
Equations (\ref{aeq}) imply the existence of $b(t,x,v,u_x)$
related to $a$ by the equations
\begin{equation}\label{btoa}
  b_t = - \varepsilon a_{u_x},\quad b_x = \varepsilon a_v,\quad
  b_v = - a_x,\quad b_{u_x} = a_t.
\end{equation}
As a consequence of (\ref{btoa}), the potential $b(t,x,v,u_x)$
satisfies the same equations (\ref{aeq}) as $a(t,x,v,u_x)$. In
terms of $b$, the result (\ref{Rfiord2}) becomes
\begin{gather}
\left( \begin{array}{c}
    \tilde{\varphi}_a
    \\ \tilde{\psi}_a
 \end{array}
\right) = \left(
 \begin{array}{c}
  D_x^{-1}(b_x + b_{u_x}u_{xx} + b_v v_x)
    \\ b_t + b_v Q + b_{u_x}v_x
 \end{array}
\right) = \left(
 \begin{array}{c}
  D_x^{-1}D_x[b]
    \\ b_t + b_v Q + b_{u_x}v_x
 \end{array}
\right)
 \notag
\\ = \left(
 \begin{array}{c}
  b(t,x,v,u_x)
    \\ b_t + b_v Q + b_{u_x}v_x
 \end{array}
\right) + \left(
 \begin{array}{c}
  c(y,z)
    \\ 0
 \end{array}
\right) = \left(
 \begin{array}{c}
    \varphi_b
    \\ \psi_b
 \end{array}
\right)  + \left(
 \begin{array}{c}
  \varphi_{3c}
    \\ \psi_{3c}
 \end{array}
\right),
 \label{Rfiord2f}
\end{gather}
so that, up to an arbitrary ``constant'' of integration $c(y,z)$,
the recursion acts on the space of second-order symmetries and on
solutions of linear system (\ref{aeq}).

The repeated application of the transformation (\ref{btoa}) to
$\tilde{a} = b$ takes us back to $a$: $\tilde{b} = - \varepsilon
a$, so that the action of the recursion on (\ref{Rfiord2f})
results in the formula
\begin{equation}\label{btoa2}
 \left( \begin{array}{c}
    \tilde{\tilde{\varphi}}_a
    \\[1mm] \tilde{\tilde{\psi}}_a
 \end{array}
\right) = R\left(
 \begin{array}{c}
    \tilde{\varphi_a}
    \\ \tilde{\psi_a}
 \end{array}
\right) = - \varepsilon\left(
 \begin{array}{c}
    \varphi_a
    \\ \psi_a
 \end{array}
\right) - \left(
 \begin{array}{c}
  c_zu_y - c_yu_z,\\
  c_zv_y - c_yv_z
 \end{array}
\right)  + \left(
 \begin{array}{c}
  d(y,z)
    \\ 0
 \end{array}
\right) ,
\end{equation}
 where we have used the relation (\ref{Rfi3}) and $d(y,z)$ is an arbitrary
 "constant" of integration.

The transformed characteristics of $X_6$, $(\tilde{\varphi_6},
\tilde{\psi_6})$ in (\ref{Rfi6}), can be obtained by the
transformation (\ref{btoa}) from $a(t,x,v,u_x) = \varphi_6 = xu_x
- tv$, which yields $\tilde{\varphi}_6 = b = - (u_xv + \varepsilon
xt)$.

The transformation (\ref{btoa}), being applied to mixed heavenly
system (\ref{mix2}) itself, transforms $a = v$ into $b =
\varepsilon (x - C)$, where $C$ is an arbitrary constant, so that
the transformed flow becomes
\begin{equation}\label{Rmixheav}
  \left(
\begin{array}{c}
  u_{\tilde{t}} \\ v_{\tilde{t}}
 \end{array}
\right) =
  R\left(
 \begin{array}{c}
    v
    \\ Q
 \end{array}
\right) = \left(
 \begin{array}{c}
  \varepsilon (x - C)
    \\ 0
 \end{array}
\right) + \left(
 \begin{array}{c}
  c(y,z)
    \\ 0
 \end{array}
\right) ,
\end{equation}
where $\tilde{t}$ is the time transformed by the recursion. We
note that the action of $R$ on $(\varphi_c, \psi_c)$ in
(\ref{Rficf}) is a particular case of the action of $R$ on
second-order symmetries in (\ref{Rfiord2f}).

\subsection{Hamiltonian structure of symmetry flows and\\
conservation laws}
 \label{sec-Hamiltonians}

Hamiltonian operators provide the natural link between commuting
symmetries in evolution form \cite{olv} and conservation laws
(integrals of motion) in involution with respect to Poisson
brackets. We write Lie equations for symmetries with the
two-component characteristics $(\varphi_i, \psi_i)$ in the
Hamiltonian form
\begin{equation}
 \left(
 \begin{array}{c}
  u_{\tau_i} \\ v_{\tau_i}
 \end{array}
 \right) =
 \left(
 \begin{array}{c}
 \varphi_i \\ \psi_i
 \end{array}
 \right) =  J_0 \left(
 \begin{array}{c}
 \delta_u H^i \\ \delta_v H^i
 \end{array}
 \right),
 \label{flow}
 \end{equation}
where the symmetry group parameter $\tau_i$ plays the role of
$ith$ time for the flow (\ref{flow}) and ${\cal
H}^i=\iiint_{-\infty}^{+\infty} H^i dx dy dz$ is an integral of
the motion along the flow (\ref{mix2}), with the conserved density
$H^i$. The second equality in (\ref{flow}) is the Hamiltonian form
of Noether's theorem that gives a relation between symmetries and
integrals. We determine conserved densities $H^i$, corresponding
to known symmetry characteristics $(\varphi_i, \psi_i)$, by
inverting the relation (\ref{flow}) in the form of \textit{inverse
Noether theorem}
\begin{equation}
\left(
\begin{array}{c}
\delta_u H^i \\ \delta_v H^i
\end{array}
\right) = K \left(
\begin{array}{c}
\varphi_i \\ \psi_i
\end{array}
\right) = \left(
\begin{array}{c}
u_{xy}\varphi_{i\,z} - u_{xz}\varphi_{i\,y} + 2v_x\varphi_{i\,x} +
v_{xx}\varphi_i
- u_{xx}\psi_i \\
u_{xx}\varphi_i
\end{array} \right),
 \label{noether}
\end{equation}
where symplectic operator $K = J_0^{-1}$ is defined in (\ref{K}).

By using (\ref{noether}), we find Hamiltonians for the first four
symmetries from the list (\ref{point})
\begin{eqnarray}
 & & X_1:\ H^1_1 = \left(tv - \frac{1}{2}\,u\right)u_{xx} \iff H^1 = \left(tv - xu_x\right)u_{xx},
  \nonumber
 \\ & & X_2:\ H^2 = vu_xu_{xx},\quad X_3^{a}:\ H^3_{a(y,z)} = a(y,z)vu_{xx} + \frac{u}{2}\,
 (a_zu_{xy} - a_yu_{xz}),
 \nonumber
 \\ & & X_4^{a}:\ H^4_{a(y,z)} = (a_zu_y - a_yu_z)\left\{vu_{xx} +
 \frac{1}{3}(u_yu_{xz} - u_zu_{xy})\right\},
\label{Hamsym}
\end{eqnarray}
while for the symmetry $X_5$ Hamiltonian does not exist and hence
this is not a variational symmetry. Equivalent Hamiltonian
densities $H^1_1$ and $H^1$ in (\ref{Hamsym}) differ by a total
$x$-derivative, which vanishes in the Hamiltonian. Hamiltonians
for symmetries $X_c$ and $X_6$ are obtained below as
specializations of Hamiltonians for second-order symmetries.

Lie equations of second-order symmetries (\ref{Lieord2}) of system
(\ref{mix2}) admit Hamiltonian form (\ref{flow}) with the
Hamiltonian density
\begin{equation}\label{Hamil2}
  H^a = A(t,x,v,u_x) u_{xx} - \gamma(t)u,
\end{equation}
where $\gamma(t)$ is an arbitrary function and $A$ is defined in
terms of the function $a(t,x,v,u_x)$ in (\ref{Lieord2}) by the
relation $A_v(t,x,v,u_x) = a(t,x,v,u_x)$. Here $A$ should satisfy
the equations
\begin{gather}
A_{tx} - \varepsilon A_{vu_x} = \alpha(t,x,u_x),\qquad  A_{tv} +
A_{xu_x} = 0,\notag
 \\ A_{xx} + \varepsilon A_{vv} = \gamma(t),\qquad \phantom{A_{tt} + \varepsilon}
 A_{tt} + \varepsilon A_{u_xu_x} = \delta(t,x,u_x)
 \label{Aeq}
\end{gather}
as a consequence of the corresponding equations (\ref{aeq}) for
the function $a$ with restrictions on some ``constants'' of
integration, that follow from (\ref{noether}). We note that at $A
= v^2/2$ we have $a = v$ and, as a consequence of (\ref{Aeq}),
$\gamma = \varepsilon$ and the Hamiltonian (\ref{Hamil2}) reduces
to the Hamiltonian density $H_1$ defined in (\ref{H1}), while
second-order Lie equations (\ref{Lieord2}) reduce to mixed
heavenly system (\ref{mix2}). Therefore, the mixed heavenly system
is embedded into the hierarchy of second-order flows commuting
with it.

We show now that Hamiltonians for the symmetry flows generated by
$X_c$, $X_1$, $X_2$ and $X_6$ can be obtained as particular cases
of Hamiltonians (\ref{Hamil2}) for second-order symmetries. The
resulting Hamiltonians below are simplified by eliminating the
terms which are total $x$-derivatives.

Hamiltonian $H^1$ of the flow generated by $X_1$, given in
(\ref{Hamsym}), is a particular case of $H^a$ in (\ref{Hamil2}) at
$a=A_v=t$ with $A = tv - xu_x$, that satisfies equations
(\ref{Aeq}) with $\alpha = \gamma = \delta = 0$. Hamiltonian $H^2$
in (\ref{Hamsym}) of the flow generated by $X_2$ is a particular
case of $H^a$ in (\ref{Hamil2}) at $a=A_v=u_x$ with $A = vu_x$,
that satisfies equations (\ref{Aeq}) with $\alpha = - \varepsilon$
and $\gamma = \delta = 0$.

Hamiltonian $H^6$ of the flow generated by $X_6$ in (\ref{point})
is obtained from the Hamiltonian (\ref{Hamil2}) of the
second-order flow by setting $a = A_v = \varphi_6 = xu_x - tv$
which, on account of equations (\ref{Aeq}) for $A$, implies
$\gamma = 0$ and specifies $A$ and $H$ in (\ref{Hamil2}) as
\begin{equation}\label{H_6}
  A_6 = \frac{1}{2}\,t(\varepsilon x^2 - v^2) + x u_xv, \qquad
  H^6 = A_6(t,x,v,u_x)u_{xx}.
\end{equation}
Here again we have dropped the terms which are total
$x$-derivatives which resulted in the simplification of $H^6$ and
$A_6$ in (\ref{H_6}), so that the equations (\ref{Aeq}) for $A$
are now satisfied with $\alpha = \gamma = \delta = 0$.

Hamiltonian of the flow generated by $X_c$ in (\ref{point}) is
obtained from $H^a$ in (\ref{Hamil2}) by setting $a = A_v =
c(x,v)$:
\begin{equation}\label{H_c}
  H^c = d(x,v)u_{xx} ,\qquad \textrm{where}\quad  d_{xx}(x,v) + \varepsilon d_{vv}(x,v) = 0
\end{equation}
and the function $d(x,v)$ is defined in terms of $c(x,v)$ by the
relation $d_v(x,v) = c(x,v)$. Here $A = d(x,v)$ again satisfies
equations (\ref{Aeq}) with $\alpha = \gamma = \delta = 0$.

\subsection{Action of recursion operator on Hamiltonians of
symmetry flows}
 \label{sec-RH}

Transformation (\ref{Rfi1fin}) of the symmetry $(\varphi_1,
\psi_1)$ implies the following transformation of Hamiltonian $H^1$
\begin{equation}\label{H1tr}
  \tilde{H}^1 = H^2 + H^3_{c(y,z)} = vu_xu_{xx} + c(y,z)v u_{xx}
  + \frac{u}{2}\,(c_zu_{xy} - c_yu_{xz}),
\end{equation}
 where $H^3_{c(y,z)}$ is determined in (\ref{Hamsym}) with $a(y,z)$
 replaced by $c(y,z)$.

The action of $R$ on the symmetry $(\varphi_2, \psi_2)$ in
(\ref{Rfi2fin}) induces the transformation of $H^2$
\begin{equation}\label{H2tr}
  \tilde{H}^2 = - \varepsilon H^1 + H^3_{c(y,z)} = - \varepsilon
  (tv - xu_x)u_{xx} + H^3_{c(y,z)} .
\end{equation}
Recursion (\ref{Rfi3}) for the symmetry $(\varphi_3, \psi_3)$ in
terms of its Hamiltonian $H^3_{a(y,z)}$ becomes
\begin{equation}\label{H3tr}
  \tilde{H}^3_{a(y,z)} = - H^4_{a(y,z)} + H^3_{c(y,z)},
\end{equation}
where $H^4_{a(y,z)}$ is determined in (\ref{Hamsym}). For the
symmetry $(\varphi_4, \psi_4)$, transformed by $R$ in
(\ref{Rfi4}), the resulting nonlocal symmetry
$(\tilde{\varphi}_4,\tilde{\psi}_4)$ has a nonlocal Hamiltonian in
the representation (\ref{flow}) with Hamiltonian operator $J_0$.
In the next section, we will show that the corresponding nonlocal
symmetry flow has local Hamiltonian density $H^4$ with respect to
the Hamiltonian operator $J_1 = RJ_0$, as will follow from
(\ref{high}).

A recursion for Hamiltonian $H^6$ of the symmetry flow generated
by $X_6$ will be more convenient to consider at the end of this
section as a particular case of a recursion for Hamiltonians
(\ref{Hamil2}) of second-order symmetries (\ref{order2}).

Recursion (\ref{Rficf}) for the symmetry $(\varphi_{c(x,v)},
\psi_{c(x,v)})$ implies the following transformation of
Hamiltonian $H^c$ of this flow defined in (\ref{H_c})
\begin{equation}\label{H^c}
  \tilde{H}^c = - H^g + H^3_{a(y,z)} = - g(x,v)u_{xx} + a(y,z)v u_{xx}
  + \frac{u}{2}\,(a_zu_{xy} - a_yu_{xz}),
\end{equation}
where $a(y,z)$ is an arbitrary function, $g(x,v)$ satisfies the
equations $g_{xx} + \varepsilon g_{vv} = 0$ and $g_v(x,v) =
f(x,v)$, with $f(x,v)$ being determined in terms of $c(x,v)$ by
the relations $f_v = c_x,\, f_x = - \varepsilon c_v$.

Recursion formula (\ref{Rfiord2f}) for second-order symmetries
implies the following recursion for their Hamiltonians:
\begin{eqnarray}
 & & \tilde{H}^a = H^b + H^3_{c(y,z)} = \big[B(t,x,v,u_x) +
c(y,z)v\big]u_{xx} \nonumber
 \\ & & \qquad\qquad\qquad\qquad\quad \mbox{} + \frac{u}{2}\,(c_zu_{xy} - c_yu_{xz}
- 2\gamma(t)),
 \label{recurs2order}
\end{eqnarray}
where $B_v = b(t,x,v,u_x)$, $b$ is related to $a(t,x,v,u_x)$ by
equations (\ref{btoa}) and $b$ satisfies the same equations
(\ref{aeq}) as $a(t,x,v,u_x)$.

For the symmetry flow $(\tilde{\varphi}_6, \tilde{\psi}_6)$ in
(\ref{Rfi6}), obtained by the action of $R$ on the flow generated
by $X_6$, we take for $b = B_v$ in the Hamiltonian
(\ref{recurs2order}) the solution $b =
 - (u_xv + \varepsilon xt)$ of equations
(\ref{btoa}) with $a = xu_x - tv$. The resulting Hamiltonian
$\tilde{H}^6$ for the symmetry flow (\ref{Rfi6}) reads
\begin{equation}\label{H_tilde6}
  \tilde{H}^6 = \tilde{A}_6(t,x,v,u_x)u_{xx} + H^3_{c(y,z)},\qquad
  \tilde{A}_6 = \frac{1}{2}\,\big(\varepsilon x^2 - v^2\big)u_x
   - \varepsilon x t v,
\end{equation}
where equations (\ref{Aeq}) for $\tilde{A}_6 = B$ are satisfied at
$\gamma = \alpha = \delta = 0$.

For mixed heavenly system given in Hamiltonian form
(\ref{Hamilton}) with Hamiltonian $H_1$, defined in (\ref{H1}), we
apply transformation (\ref{btoa}) to $a=v$ to get $b = B_v =
\varepsilon (x-c)$, where $c$ comes from a constant of
integration. Then we determine $B$ in (\ref{recurs2order}),
solving (\ref{Aeq}) with $A\mapsto B$, to obtain the transformed
$H_1$ in the form
\begin{equation}\label{H1tilde}
  \tilde{H}_1 = \varepsilon (x - c)vu_{xx} + H^3_{c(y,z)} =
  - \varepsilon H_0 + H^3_{c(y,z)},
\end{equation}
 where $H_0$ is defined in (\ref{H0}).
The second application of the same transformation to
(\ref{H1tilde}) takes us back to $H_1$ (modulo the transformed
``constant'' of integration $\tilde{H}^3_{c(y,z)}$ determined by
(\ref{H3tr})):
\begin{equation}\label{H1tild2}
 \tilde{\tilde{H}}_1 = \frac{1}{2}(x^2 - \varepsilon v^2)u_{xx} +
\tilde{H}^3_{c(y,z)} = - \varepsilon H_1 + \tilde{H}^3_{c(y,z)}
\end{equation}
The last equation in (\ref{H1tild2}) is obvious from the
alternative formula for $H_1$ in (\ref{H1}).

For the second Hamiltonian density $H_0=(c-x)vu_{xx}$, the
transformed Hamiltonian coincides with $H_1$: $\tilde{H}_0 = H_1$.
This is obvious if one compares bi-Hamiltonian representation
(\ref{biHam}) of the mixed heavenly system with the second formula
in (\ref{R+delH}) applied to $H = H_0$.

The transformation inverse to (\ref{recurs2order}):
$\tilde{H}_{-1} = H_0$ (with $a$ and $b$ interchanged and
$b=B_v=c-x$), yields
\begin{equation}\label{H0tild}
  H_{-1} = \frac{1}{2}(x^2 - \varepsilon
  v^2)u_{xx} + H^3_{c(y,z)} \iff H_{-1} = - \varepsilon H_1 +
  H^3_{c(y,z)},
\end{equation}
where equivalent Hamiltonian is obtained by dropping a total
$x$-derivative.

All the flows for the variational symmetries generated by $X_1$,
$X_2$, $X_3$, $X_4$, $X_6$, $X_{c(x,v)}$, $X_{a(t,x,v,u_x)}$ and
$\tilde{X}_1, \tilde{X}_2, \tilde{X}_3, \tilde{X}_6, \tilde{X}_c,
\tilde{X}_a$, where $\tilde{X}_i = RX_i$, have the Hamiltonian
form (\ref{flow}) with the local Hamiltonian densities presented
above, while the nonlocal symmetry generated by $\tilde{X}_4$ has
the Hamiltonian form with the local Hamiltonian $H^4$
\begin{equation}
 \left(
 \begin{array}{c}
  u_{\tilde{\tau}_4} \\ v_{\tilde{\tau}_4}
 \end{array}
 \right) =
 \left(
 \begin{array}{c}
 \tilde{\varphi}_4 \\ \tilde{\psi}_4
 \end{array}
 \right) =  J_1 \left(
 \begin{array}{c}
 \delta_u H^4 \\ \delta_v H^4
 \end{array}
 \right)
 \label{flowtild4}
 \end{equation}
with respect to Hamiltonian operator $J_1$, as will be shown in
(\ref{high}).

All Hamiltonians of symmetry flows are integrals of mixed heavenly
system (\ref{mix2}).

\section{Hierarchy and bi-Hamiltonian representations for symmetry flows
of mixed heavenly system}
 \setcounter{equation}{0}
 \label{sec-highflows}

We know from the work of B. Fuchssteiner and A. S. Fokas \cite{ff}
(see also the survey \cite{sheftel} and references therein) that
if a recursion operator has a factorized form, as in our case $R =
J_1J_0^{-1}\equiv J_1K$, and the factors $J_0$ and $J_1$ are
compatible Hamiltonian operators, then $R$ is a hereditary
(Nijenhuis) recursion operator. The skew symmetry of Hamiltonian
operators $J_1^\dag = - J_1$ and $J_0^\dag = - J_0$ implies
$R^\dag = J_0^{-1}J_1 = KJ_1$, so that $RJ_0 = J_1 = J_0R^\dag$.
Note that in this section, in contrast to sections \ref{sec-Rsym}
and \ref{sec-RH}, we have $D_x^{-1} = \int_{-\infty}^x\,dx'$ (for
functions vanishing rapidly at $-\infty$) in the definition
(\ref{R}) of $R$, same as in the definition (\ref{J1}) of $J_1$,
so that $D_x^{-1}D_x = 1$.

Now, the Hermitian conjugate hereditary recursion operator
\begin{equation}\label{R^+}
R^\dag = \left(
\begin{array}{cc}
   (-D_xv_x + u_{xz}D_y - u_{xy}D_z)D_x^{-1} & D_xQ - v_zD_y + v_yD_z
\\                          - u_{xx}D_x^{-1} &        v_x
\end{array}
\right),
\end{equation}
acting on the vector of variational derivatives of an integral of
the flow, again yields a vector of variational derivatives of some
integral of this flow \cite{ff}. Therefore, acting with the
recursion operator on a variational symmetry flow
\begin{equation}\label{symflow}
\left(
\begin{array}{c}
u_\tau \\ v_\tau
\end{array}
\right) =  J_0 \left(
\begin{array}{c}
\delta_u H \\ \delta_v H
\end{array}
\right),
\end{equation}
 we obtain
\begin{equation}\label{Rflow}
  R\left(
\begin{array}{c}
u_\tau \\ v_\tau
\end{array}
\right) = \left(
\begin{array}{c}
u_{\tilde{\tau}} \\ v_{\tilde{\tau}}
\end{array}
\right) = J_1\left(
\begin{array}{c}
\delta_u H \\ \delta_v H
\end{array}
\right) = J_0R^\dag \left(
\begin{array}{c}
\delta_u H \\ \delta_v H
\end{array}
\right) = J_0\left(
\begin{array}{c}
\delta_u \tilde{H} \\ \delta_v \tilde{H}
\end{array}
\right),
\end{equation}
where $\tilde{\tau}$ and $\tilde{H}$ are the group parameter
(``time'') and Hamiltonian of the transformed symmetry obtained by
the action of $R$ on (\ref{symflow}). The transformed flow
(\ref{Rflow}) and its Hamiltonian $\tilde{H}$ have been determined
in subsections \ref{sec-Rsym} and \ref{sec-RH}, respectively,
where we now have to skip all arbitrary ``constants'' of
integration $c(y,z)$ and $a(y,z)$ for transformed symmetries
$(\tilde{\varphi},\tilde{\psi})^T$ and transformed Hamiltonians
$\tilde{H}$ due to the restricted definition of $R$ given at the
beginning of this section. Thus, an alternative (to subsection
\ref{sec-RH}) way of transforming a Hamiltonian by the recursion
operator is to act by the Hermitian conjugate recursion operator
on the vector of variational derivatives of this Hamiltonian
\begin{equation}\label{R+delH}
R^\dag \left(
\begin{array}{c}
\delta_u H \\ \delta_v H
\end{array}
\right) = \left(
\begin{array}{c}
\delta_u \tilde{H} \\ \delta_v \tilde{H}
\end{array}
\right)\quad \Longrightarrow \quad J_1\left(
\begin{array}{c}
\delta_u H \\ \delta_v H
\end{array}
\right) =  J_0\left(
\begin{array}{c}
\delta_u \tilde{H} \\ \delta_v \tilde{H}
\end{array}
\right),
\end{equation}
where both relations follows from (\ref{Rflow}). Similarly,\\
$J_2\left(
\begin{array}{c}
\delta_u H \\ \delta_v H
\end{array}
\right) =  J_1\left(
\begin{array}{c}
\delta_u \tilde{H} \\ \delta_v \tilde{H}
\end{array}
\right)$ and so on.

We now use these remarks for constructing hierarchies of mixed
heavenly system (\ref{mix2}) and its symmetry flows, together with
bi-Hamiltonian representations of the symmetry flows. We start by
applying $R$ to this system in Hamiltonian form (\ref{Hamilton})
\begin{equation}
\left(
\begin{array}{c}
u_{t_1} \\ v_{t_1}
\end{array}
\right) =  J_1 \left(
\begin{array}{c}
\delta_u H_1 \\ \delta_v H_1
\end{array}
\right) = J_0 \left(
\begin{array}{c}
\delta_u \tilde{H_1} \\ \delta_v \tilde{H_1}
\end{array}
\right) = \varepsilon
 \left(
\begin{array}{c}
   x - C \\ 0
 \end{array}
\right),
 \label{J1H1}
\end{equation}
where we have used the result (\ref{Rmixheav}), $C$ is an
arbitrary constant, $t_1 = \tilde{t}$ is the parameter of the
group transformed by the recursion and $\tilde{H_1} = -
\varepsilon H_0$ according to (\ref{H1tilde}) with $H^3_{a(y,z)}$
skipped, where $H_0$ is defined in (\ref{H0}).

Applying again the recursion operator to the Hamiltonian system
(\ref{J1H1}) we obtain
\begin{equation}
\left(
\begin{array}{c}
u_{t_2} \\ v_{t_2}
\end{array}
\right) =
  J_1 \left(
  \begin{array}{c}
  \delta_u\tilde{H_1} \\ \delta_v\tilde{H_1}
  \end{array}
  \right) = J_0
  \left(
  \begin{array}{c}
  \delta_u\tilde{\tilde{H}}_1 \\ \delta_v\tilde{\tilde{H}}_1
  \end{array}
  \right)
  = - \varepsilon J_0\left(
\begin{array}{c}
\delta_u H_1 \\ \delta_v H_1
\end{array}
\right) = - \varepsilon \left(
\begin{array}{c}
v \\ Q
\end{array}
\right),
 \label{R+}
\end{equation}
where $t_2 = \tilde{t_1}$ and the transformation (\ref{H1tild2})
of Hamiltonian $\tilde{H}_1$ was used: $\tilde{\tilde{H}}_1
 = - \varepsilon H_1$. This result shows that we are
back to the mixed heavenly system and so further applications of
the recursion operator will not generate an infinite hierarchy. We
also note that by applying $J_0$ to $H_0$ we will not get anything
new:
\begin{equation}
\left(
\begin{array}{c}
u_{t_0} \\ v_{t_0}
\end{array}
\right) =  J_0 \left(
\begin{array}{c}
\delta_u H_0 \\ \delta_v H_0
\end{array}
\right) =
 \left(
\begin{array}{c}
  C - x \\ 0
 \end{array}
\right).
 \label{J0H0}
\end{equation}

We now apply $R$ to Hamiltonian symmetry flows from (\ref{point}),
that commute with the mixed heavenly flow (\ref{mix2}), to obtain
the following results:
\begin{equation}
\left(
\begin{array}{c}
u_{\tilde{t^1}} \\ v_{\tilde{t^1}}
\end{array}
\right) =
    J_1 \left(
  \begin{array}{c}
  \delta_uH^1 \\ \delta_vH^1
  \end{array}
  \right) = J_0\left(
  \begin{array}{c}
  \delta_u\tilde{H^1} \\ \delta_v\tilde{H^1}
  \end{array}
  \right) = J_0 \left(
  \begin{array}{c}
  \delta_u H^2 \\
  \delta_v H^2
  \end{array}
  \right) = \left(
  \begin{array}{c}
  u_x \\ v_x
  \end{array}
  \right),
  \label{RH1}
\end{equation}
\begin{equation}
\left(
\begin{array}{c}
u_{\tilde{t^2}} \\ v_{\tilde{t^2}}
\end{array}
\right) =
    J_1 \left(
  \begin{array}{c}
  \delta_u H^2 \\
  \delta_v H^2
  \end{array}
  \right) = J_0\left(
  \begin{array}{c}
  \delta_u \tilde{H^2} \\
  \delta_v \tilde{H^2}
  \end{array}
  \right) = - \varepsilon J_0 \left(
  \begin{array}{c}
  \delta_u H^1 \\
  \delta_v H^1
  \end{array}
  \right)
    = - \varepsilon\left(
  \begin{array}{c}
  t \\ 1
  \end{array}
  \right),
 \label{RH2}
\end{equation}
\begin{gather}
\left(
\begin{array}{c}
u_{\tilde{t^3}} \\ v_{\tilde{t^3}}
\end{array}
\right) = J_1  \left(
  \begin{array}{c}
  \delta_u H^3_{a(y,z)} \\
  \delta_v H^3_{a(y,z)}
  \end{array}
  \right) = J_0 \left(
  \begin{array}{c}
  \delta_u \tilde{H^3}_{a(y,z)} \\
  \delta_v \tilde{H^3}_{a(y,z)}
  \end{array}
  \right)
  = - J_0\left(
  \begin{array}{c}
  \delta_u H^4_a \\ \delta_v H^4_a
  \end{array}
  \right) \notag
  \\ \qquad\qquad\qquad\qquad\qquad\qquad\qquad\qquad\qquad\qquad\qquad =  - \left(
  \begin{array}{c}
  \varphi_{4a} \\ \psi_{4a}
  \end{array}
  \right).
 \label{RH3}
\end{gather}
By applying $R$ to the flow of $H^4$ we obtain the first nonlocal
Hamiltonian flow in the infinite hierarchy of symmetries of mixed
heavenly system:
\begin{equation}
 \left(
 \begin{array}{c}
  u_{\tilde{t^4}} \\ v_{\tilde{t^4}}
 \end{array}
 \right) =
 J_1 \left(
 \begin{array}{c}
 \delta_u H^4_a \\ \delta_v H^4_a
 \end{array}
 \right) = J_0 \left(
  \begin{array}{c}
  \delta_u \tilde{H^4}_a \\ \delta_v \tilde{H^4}_a
  \end{array}
  \right) = \left(
  \begin{array}{c}
  \tilde{\varphi}_{4a} \\ \tilde{\psi}_{4a}
  \end{array}
  \right),
  \label{high}
\end{equation}
where $\tilde{\varphi}_{4a}$ and $\tilde{\psi}_{4a}$ are defined
in (\ref{Rfi4}). The next higher flow of this hierarchy can be
obtained by applying $J_2$ to $(\delta_u H^4, \; \delta_v H^4)$
and so on.

Finally, we consider the action of the recursion operator on the
general flow of second-order symmetries (\ref{order2}) with the
Hamiltonian (\ref{Hamil2})
\begin{equation}\label{J0Ha}
 \left(
 \begin{array}{c}
  u_t \\ v_t
 \end{array}
 \right) =
 J_0 \left(
 \begin{array}{c}
 \delta_u H^a \\ \delta_v H^a
 \end{array}
 \right) =
 \left(
 \begin{array}{c}
   a(t,x,v,u_x) \\ a_t + a_vQ + a_{u_x}v_x
\end{array}
 \right) .
\end{equation}
 Transformation (\ref{recurs2order}) of Hamiltonian
(\ref{Hamil2}) under the action of $R$, $\tilde{H}^a = H^b$,
implies the following transformation of the flow
\begin{gather}
 \left(
 \begin{array}{c}
  u_{\tilde{t}} \\ v_{\tilde{t}}
 \end{array}
 \right) =
 J_0 \left(
 \begin{array}{c}
 \delta_u \tilde{H^a} \\ \delta_v \tilde{H^a}
 \end{array}
 \right) =
 J_0 \left(
 \begin{array}{c}
 \delta_u H^b \\ \delta_v H^b
 \end{array}
 \right) =
  \left(
 \begin{array}{c}
   b(t,x,v,u_x) \\ b_t + b_vQ + b_{u_x}v_x
\end{array}
 \right),
\label{RHa}
\end{gather}
where $b = \tilde{a}$ is related to $a$ by transformation
(\ref{btoa}), with $a$ and $b$ satisfying same equations
(\ref{aeq}). Repeated application of $R$ to (\ref{recurs2order})
yields $\tilde{\tilde{H^a}} = \tilde{H^b} = - \varepsilon H^a$,
because $\tilde{b}= - \varepsilon a$, and therefore
\begin{equation}
 \left(
 \begin{array}{c}
  u_{\,\tilde{\tilde{t}}} \\ v_{\,\tilde{\tilde{t}}}
 \end{array}
 \right) =
 J_0 \left(
 \begin{array}{c}
 \delta_u \tilde{\tilde{H^a}} \\ \delta_v \tilde{\tilde{H^a}}
 \end{array}
 \right) =  - \varepsilon
 J_0 \left(
 \begin{array}{c}
 \delta_u H^a \\ \delta_v H^a
 \end{array}
 \right)
 =  - \varepsilon  \left(
 \begin{array}{c}
   a(t,x,v,u_x) \\ a_t + a_vQ + a_{u_x}v_x
\end{array}
 \right),
\label{R2Ha}
\end{equation}
 so that we are back to original flow (\ref{J0Ha}). This is quite
similar to the behavior of the mixed heavenly system which is a
very particular case of this general flow.

By definition of a hereditary recursion operator, $R$ generates an
Abelian symmetry algebra out of commuting symmetries. Since
$\left[X_3^{a(y,z)},X_4^{a(y,z)}\right] = 0$ this implies
$\left[\tilde{X}_3^{a(y,z)},\tilde{X}_4^{a(y,z)}\right] = 0$,
where $\tilde{X}$ is the symmetry generator obtained from $X$ by
the action of $R$. Now, $\tilde{X}_3^{a(y,z)} = - X_4^{a(y,z)}$
for vanishing constant of integration $c(y,z)$ and therefore
$\left[X_4^{a(y,z)},\tilde{X}_4^{a(y,z)}\right] = 0$. A
straightforward calculation shows that the flows
$(u_{t^4},v_{t^4})^T = (\varphi_{4a},\psi_{4a})^T$ and
$(u_{\tilde{t^4}},v_{\tilde{t^4}})^T =
(\tilde{\varphi}_{4a},\tilde{\psi}_{4a})^T$ indeed commute.

Repeating this reasoning for powers of $R$ applied to the last
result, we see that the hierarchy of symmetries generated by $R$
from $X_3^{a(y,z)}$ consists of commuting flows.

\section{Two-component form and Lagrangian of\\ Husain heavenly equation}
\setcounter{equation}{0}
 \label{sec-husain}

Husain equation ($\varepsilon = + 1$ in \cite{husain})
\begin{equation}\label{husain}
  v_{ty}v_{pz} - v_{tz}v_{py} + v_{tt} + \varepsilon v_{pp} = 0
\end{equation}
can be obtained from the mixed heavenly equation (\ref{mix}) by
the partial Legendre transformation in $x$
\begin{equation}\label{legmixhus}
 x = - v_p,\quad u = v - pv_p,\qquad p = u_x,\quad v(t,p,y,z) = u - xu_x .
\end{equation}
Up to a change of notation of variables, Eq. (\ref{husain}) with
$\varepsilon = \pm 1$ is a particular case (\ref{husain_u}) of our
general equation (\ref{Ffinal}) admitting partner symmetries.

To discover its bi-Hamiltonian structure, we have to consider
Husain equation in a two-component form, without using Ashtekar's
Hamiltonian formulation of general relativity \cite{ashtek,AJS},
which was a starting point in the paper \cite{husain} by V.
Husain.

We start with the Lagrangian of Eq. (\ref{husain}) in a
one-component form
\begin{equation}\label{Lagrange}
  L = \frac{1}{2}\,(v_t^2 + \varepsilon v_p^2) + \frac{1}{3}\, v_t
  (v_yv_{pz} - v_zv_{py}) ,
\end{equation}
which for $\varepsilon = + 1$ is equivalent to the one given in
\cite{PlebPrza}.

In a two-component evolution form, Eq. (\ref{husain}) becomes
\begin{equation}\label{2comp}
  \left\{
  \begin{array}{l}
    v_t = q
 \\ q_t = q_zv_{py} - q_yv_{pz} - \varepsilon v_{pp} \,.
  \end{array} \right.
\end{equation}
We shall call Eq. (\ref{2comp}) \textit{Husain system.}

The determining equation and recursion for symmetries of Husain
equation in a one-component form could easily be obtained by
Legendre transformation (\ref{legmixhus}) from corresponding
formulas for mixed heavenly equation and then converted to a
two-component form, resulting in a Lax representation of Husain
system (\ref{2comp}). However, the main objects of the Hamiltonian
formulation, like Lagrangian, symplectic two-form, Hamiltonian
operators and Hamiltonian densities cannot be obtained that way.
Therefore, we have to undertake an independent study of Husain
system along the same lines as we have done before for mixed
heavenly system. Lagrangian for Husain system (\ref{2comp}) reads
\begin{equation}\label{L2comp}
  L = \frac{1}{2}\,\Big(2v_tq - q^2 + \varepsilon v_p^2\Big) +
  \frac{1}{3}\,v_t(v_yv_{pz} - v_zv_{py}).
\end{equation}
Note that the form of the Lagrangian (\ref{L2comp}) is not
uniquely defined by one-component Lagrangian (\ref{Lagrange}) and
it requires some skill in order to arrive at the form of
Lagrangian that will be suitable for a Hamiltonian form of Husain
system (\ref{2comp}).

\section{Symplectic and Hamiltonian structure of Husain
system} \setcounter{equation}{0}
 \label{sec-HusHam}

Lagrangian (\ref{L2comp}) yields the canonical momenta
\begin{equation}\label{canon}
  \pi_v = q + \frac{1}{3}\,(v_yv_{pz} - v_zv_{py}),\quad \pi_q = 0,
\end{equation}
that cannot be inverted for the velocities $u_t$ and $v_t$, and,
therefore, Lagrangian (\ref{L2comp}) is  degenerate. According
Dirac's theory of constraints \cite{dirac}, we treat the
definitions (\ref{canon}) as constraints of the second class
\begin{equation}\label{constrhus}
  \phi_v = \pi_v - q - \frac{1}{3}\,(v_yv_{pz} - v_zv_{py})
  = 0, \quad \phi_q = \pi_q = 0 .
\end{equation}
Poisson brackets of the constraints yield the skew-symmetric
symplectic matrix operator
\begin{equation}\label{Khus}
K =  \left( \begin{array}{cc}
 v_{py}D_z - v_{pz}D_y &\hspace*{2mm} - 1 \\
  1 &\hspace*{5mm} 0
\end{array}  \right) .
\end{equation}
The corresponding symplectic two-form $\Omega =
\int\limits_{V}\omega dx dy dz$ of the density
\begin{equation}
   \label{sympform2}
 \omega = \frac{1}{2} \left(v_{py} dv \wedge dv_z - v_{pz} d v\wedge d v_y
 - 2 dv \wedge dq \right)
\end{equation}
is closed since the exterior differential of (\ref{sympform2}) is
a total divergence (similar to (\ref{domeg}))
\begin{equation*}
  d\omega = \frac{1}{2}\,d v_z\wedge d v_p \wedge d v_y ,
\end{equation*}
which implies vanishing $\Omega$ at appropriate boundary
conditions. Therefore, $\Omega$ and $K$ in (\ref{Khus}) is indeed
a symplectic form and symplectic operator, respectively. Hence,
its inverse $K^{-1}$ is a Hamiltonian operator
\begin{equation}\label{J0Hus}
 J_0 = K^{-1} = \left(
\begin{array}{cc}
     0        &           1
\\ - 1        &    v_{py}D_z - v_{pz}D_y
\end{array}
\right).
\end{equation}
Indeed, it is explicitly skew-symmetric and Jacobi identity is
satisfied as a consequence of closeness of the form $\Omega$.

Hamiltonian density, corresponding to $J_0$, reads
\begin{equation}
 H_1 = \pi_v v_t + \pi_q q_t - L = \frac{1}{2}\,q^2 - \frac{\varepsilon}{2}\, v_p^2 ,
 \label{Hus1}
\end{equation}
so that Husain system (\ref{2comp}) is the Hamiltonian system
\begin{equation}
\left(
\begin{array}{c}
v_t \\ q_t
\end{array}
\right) =  J_0 \left(
\begin{array}{c}
\delta_v H_1 \\ \delta_q H_1
\end{array}
\right),
 \label{HamilHus}
\end{equation}
where $\delta_v$ and $\delta_q$ are Euler-Lagrange operators
\cite{olv} with respect to $v$ and $q$.

\section{Recursion operator and Lax representation for Husain system}
\setcounter{equation}{0}
 \label{sec-recurs}

Recursion relation (\ref{Husrecurs_u}) for symmetries of the
equation (\ref{husain}), with the change of notation $u\mapsto v$
and $x\mapsto p$, becomes
\begin{gather}
\tilde{\varphi}_t = v_{tz}\varphi_y - v_{ty}\varphi_z -
\varepsilon \varphi_p + \omega_0\varphi_t ,\notag
\\ \tilde{\varphi}_p = v_{pz}\varphi_y -
v_{py}\varphi_z + \varphi_t + \omega_0\varphi_p .
 \label{Husrecurs}
\end{gather}

As before, we introduce two-component symmetry characteristics of
Husain system determined by the Lie equations (with independent
variables not transformed by the group)
\begin{equation}\label{LieHus}
 \left( \begin{array}{c}
 v_\tau \\ q_\tau
 \end{array}
\right)
 = \left(
 \begin{array}{c}
    \varphi
    \\ \psi
 \end{array}
\right) \equiv  \Phi ,
\end{equation}
where $\tau$ is the group parameter. The symmetry condition is the
linear matrix equation, the compatibility condition of Lie
equations (\ref{LieHus}) and equations (\ref{2comp})
\begin{equation}\label{symHus}
 \left\{
 \begin{array}{c}
v_{t\tau} - v_{\tau t} = 0 \\
q_{t\tau} - q_{\tau t} = 0
 \end{array} \right.
\quad \iff \quad \hat{A}(\Phi) = 0 ,
\end{equation}
where $\hat{A}$ is Frech\'et derivative of the flow (\ref{2comp})
\begin{equation}
 \hat{A} =  \left( \begin{array}{cc} D_t & - 1
\\ \varepsilon D_p^2
 - q_z D_pD_y + q_y D_pD_z,
 & D_t + v_{pz}D_y - v_{py}D_z \end{array} \right).
 \label{AHus}
\end{equation}
Then the first row of (\ref{symHus}) yields $\varphi_t = \psi$.

Using the relations $\psi = \varphi_t$, and $\tilde{\psi} =
\tilde{\varphi_t}$ for the transformed symmetry, we rewrite the
recursion relation (\ref{Husrecurs}) with $\omega_0 = 0$ and $v_t
= q$ in the two-component form
\begin{gather}
\tilde{\varphi}_p = \big(v_{pz}D_y - v_{py}D_z\big)\varphi + \psi,
\notag
\\   \tilde{\psi} = \big(q_{z}D_y - q_yD_z - \varepsilon D_p\big)\varphi.
 \label{recursHus}
\end{gather}
In the notation $\tilde{\Phi} = \left(
 \begin{array}{c}
    \tilde{\varphi}
    \\ \tilde{\psi}
 \end{array}
\right)$, the recursion relation takes the form $\tilde{\Phi} =
R\big(\Phi\big)$, where recursion operator $R$ has the $2\times 2$
matrix form
\begin{equation}\label{RHus}
 R = \left(
  \begin{array}{lc}
 D_p^{-1}\Big(v_{pz}D_y - v_{py}D_z\Big)
 &  D_p^{-1}
 \\[2mm] q_zD_y - q_yD_z - \varepsilon D_p & 0
  \end{array}
  \right).
\end{equation}
As in the case of the mixed heavenly system, recursion relation
(\ref{recursHus}) for symmetries implies that $D_p^{-1}$ should be
understood as an indefinite integral with respect to $p$ with the
``constant'' of integration $C(y,z,t)$ determined by the
constraint $\tilde{\varphi}_t = \tilde{\psi}$ up to an arbitrary
additive term $c(y,z)$. However, in the Lax representation
(\ref{commutH}) and definition of the second Hamiltonian operator
$J_1 = RJ_0$ in (\ref{J1Hus}) we should use the restricted
definition of $D_p^{-1}$ as the definite integral
$\int_{-\infty}^p dp^{\,\prime}$ satisfying the condition
$D_p^{-1}D_p = 1$ (on assumption that all functions vanish at
$p=-\infty$).

The commutator of recursion operator $R$ and operator $\hat{A}$ of
symmetry condition (\ref{symHus}), computed without using the
equations of motion, reads
\begin{eqnarray}
 & & [R , \hat{A}] =
 \label{commutH}
 \\ & & \left[
\begin{array}{cc}
  D_p^{-1}\big[(v_t - q)_{py}D_z - (v_t - q)_{pz}D_y \big], & 0
 \\ (q_t - q_zv_{py} + q_yv_{pz} + \varepsilon v_{pp})_y D_z
 - (q_t - q_zv_{py} + q_yv_{pz} + \varepsilon v_{pp})_z D_y ,  & 0
\end{array}
 \right] .
 \nonumber
\end{eqnarray}
Thus, on solutions of the system (\ref{2comp}) operators $R$ and
$\hat{A}$ commute and therefore $R$ acting on any symmetry $\Phi$
generates again a symmetry, so that $R$ is indeed a recursion
operator. Furthermore, vanishing of the commutator (\ref{commutH})
reproduces Husain system (\ref{2comp}) and hence the operators $R$
and $\hat{A}$ form a Lax pair of the Olver-Ibragimov-Shabat type
\cite{olver,ibr}.

\section{Bi-Hamiltonian representation of Husain system}
\setcounter{equation}{0}
 \label{sec-biHamiltonHus}

By a theorem of Magri \cite{magri}, second Hamiltonian operator is
obtained by acting with recursion operator (\ref{RHus}) on
Hamiltonian operator (\ref{J0Hus})
\begin{equation}\label{J1Hus}
  J_1 = RJ_0 = \left(
  \begin{array}{cc}
    - D_p^{-1}  &  0
   \\ 0         &  q_zD_y - q_yD_z - \varepsilon D_p
  \end{array}
  \right),
\end{equation}
which is explicitly skew-symmetric. The proof of the Jacobi
identity has been performed by using P. Olver's criterion
\cite{olv} in terms of functional multivectors.

We have also made a check for compatibility of the two Hamiltonian
operators $J_0$ and $J_1$ by using P. Olver's criterion and proved
that every linear combination $aJ_0 + bJ_1$ with arbitrary
constant coefficients $a$ and $b$ satisfies the Jacobi identity,
i.e. $J_0$ and $J_1$ form a Poisson pencil (a compatible
Hamiltonian pair). The flow (\ref{2comp}) is generated by $J_1$
from the Hamiltonian density
\begin{equation}\label{H0Hus}
  H_0 = q v_p .
\end{equation}
Thus, Husain equation in two-component form (\ref{2comp}) is a
{\it bi-Hamiltonian integrable system}:
\begin{equation}\label{biHamHus}
\left(
\begin{array}{c}
v_t \\ q_t
\end{array}
\right) =  J_0 \left(
\begin{array}{c}
\delta_v H_1 \\ \delta_q H_1
\end{array}
\right) =  J_1 \left(
\begin{array}{c}
\delta_v H_0 \\ \delta_q H_0
\end{array}
\right).
\end{equation}

\section{Symmetries and conservation laws for\\ Husain system}
\setcounter{equation}{0}
 \label{sec-Husymmetries}

Using symmetry package LIEPDE in REDUCE \cite{wolf}, we have
computed all generators and two-component characteristics
$\varphi, \psi$ of point symmetries of Husain system
(\ref{2comp}):
\begin{eqnarray}
 & & X_1 = \partial_t , \quad \varphi_1 = - q,\quad
 \psi_1 = q_y v_{pz} - q_z v_{py} + \varepsilon v_{pp}
 \nonumber
\\ & & X_2 = - t\partial_t - p\partial_p + q\partial_q, \nonumber
 \\ & & \varphi_2 = t q + pv_p,\quad \psi_2 = q + pq_p
 + t(q_zv_{py} - q_yv_{pz} - \varepsilon v_{pp})
 \nonumber
  \\ & & X_3 = \partial_p, \quad \varphi_3 = - v_p,\quad
  \psi_3 = - q_p
  \label{pointHus}
  \\ & & X_4 = \frac{1}{2}\left(y\partial_y + z\partial_z\right) + v\partial_v + q\partial_q,
    \nonumber
  \\ & & \varphi_4 = v - \frac{1}{2}\left(yv_y + zv_z\right),\;
    \psi_4 = q - \frac{1}{2}\left(yq_y + zq_z\right),
  \nonumber
  \\ & & X_5^a = a(y,z)\partial_v,\quad \varphi_{5a} = a(y,z),\quad \psi_{5a} = 0
  \nonumber
  \\ & & X_6^a = a_y(y,z)\partial_z - a_z(y,z)\partial_y,\quad
  \varphi_{6a} = a_zv_y - a_yv_z,\quad \psi_{6a} = a_zq_y - a_yq_z
  \nonumber
  \\ & & X_{c(t,p)} = c_t(t,p)\partial_v + c_{tt}(t,p)\partial_q,
  \quad \varphi_c = c_t(t,p),\quad \psi_c = c_{tt}(t,p),
  \nonumber
\end{eqnarray}
where $c(t,p)$ is an arbitrary smooth solution of the equation
\begin{equation}\label{ceq}
c_{tt} + \varepsilon c_{pp} = 0
\end{equation}
and we have used equations of motion (\ref{2comp}) for eliminating
$v_t$ and $q_t$. Translations in $y,z$ and $v$ can be obtained as
simple particular cases of $X_6^a$ and $X_5^a$, respectively.

All second-order Lie-B\"{a}cklund symmetries of Husain system
modulo point symmetries have the generators
\begin{gather}
\hat{X}_f = f(t,p,q,v_p)\partial_v + \left\{ f_t + f_q(q_zv_{py} -
q_yv_{pz} - \varepsilon v_{pp}) + q_pf_{v_p}\right\}\partial_q +
\cdots , \notag \\
 \varphi_f = f(t,p,q,v_p),\quad \psi_f = f_t + f_q(q_zv_{py} -
q_yv_{pz} - \varepsilon v_{pp}) + q_pf_{v_p} ,
 \label{gen2Hus}
\end{gather}
where $f(t,p,q,v_p)$ satisfies the equations
\begin{equation}\label{eqf}
  f_{tv_p} + \varepsilon f_{pq} = 0,\quad f_{pv_p} - f_{qt} = 0,
\quad f_{qq} + \varepsilon f_{v_pv_p} = 0,\quad f_{tt} +
\varepsilon f_{pp} = 0.
\end{equation}
Lie equations corresponding to (\ref{gen2Hus}) have the form of a
second-order flow
\begin{equation}\label{order2Hus}
  v_\tau = f(t,p,q,v_p),\quad q_\tau = f_t + f_q(q_zv_{py} - q_yv_{pz}
- \varepsilon v_{pp}) + q_pf_{v_p} ,
\end{equation}
where the ``time'' $\tau$ is the group parameter. Husain system is
itself a particular case of (\ref{order2Hus}) at $f=q$ obviously
satisfying conditions (\ref{eqf}), so that symmetry generator
$X_1$ in evolutionary form is a particular case of $\hat{X}_f$.
Similarly, we note that symmetries $X_2$, $X_3$ and $X_{c(t,p)}$
are also particular cases of the second-order symmetry
$X_{f(t,p,q,v_p)}$, while the symmetries $X_4$, $X_5^{a(y,z)}$ and
$X_6^{a(y,z)}$ are not.

All second-order Lie-B\"acklund symmetries can be obtained by
taking linear combinations of generators (\ref{gen2Hus}) and point
symmetry generators $X_4$, $X_5^{a(y,z)}$ and $X_6^{a(y,z)}$ in
evolutionary form.
\begin{table}[ht]
\caption{Commutators of symmetries of Husain system.}
\rule[2mm]{13.3cm}{1pt} \\
{\begin{tabular}{lc@{\hspace{2pt}}c@{\hspace{1pt}}c@{\hspace{-.05pt}}ccc@{\hspace{-.05pt}}c@{\hspace{-.05pt}}c}
              &$X_1$&$X_2$&$X_3$ & $X_4$ & $X_5^{f(y,z)}$
              & $X_6^{b(y,z)}$ & $X_{c(t,p)}$ & $\hat{X}_{g(t,p,q,v_p)}$
              \\[2pt] \hline \\[-2pt]
 $X_1$        &  0 & 0  & 0              & 0             & 0& 0 & $X_{c_t}$ & $\hat{X}_{g_t}$
 \\[2pt]
 $X_2$        &  0 & 0  & $X_3$             & 0              & 0&0& $-X_{c^\prime}$& $-\hat{X}_{\tilde{g}}$  \\
 $X_3$ & 0 & $-X_3$  & 0             & 0  & 0  & 0 & $X_{c_p}$ & $\hat{X}_{g_p}$ \\
 $X_4 $& 0 & 0  & 0 & 0 & $\frac{1}{2}X_5^{\hat{f}(y,z)}$& $\frac{1}{2}X_6^{\hat{b}(y,z)}$ & $-X_c$ & $\hat{X}_{\breve{g}}$
 \\[2pt]  $X_5^{a(y,z)}$         & 0 & 0  & 0 & $-\frac{1}{2}X_5^{\hat{a}}$& 0 &
 $X_5^{\frac{\partial(a,b)}{\partial(y,z)}}$
 & 0 & 0 \\
 $X_6^{a(y,z)}$   & 0 & 0 & 0 & $-\frac{1}{2}X_6^{\hat{a}}$ & $-X_5^{\frac{\partial(f,a)}{\partial(y,z)}}$
  & $X_6^{\frac{\partial(a,b)}{\partial(y,z)}}$ & 0 & 0 \\
 $X_{d(t,p)}$ & $-X_{d_t}$ & $X_{d^\prime}$ & $-X_{d_p}$ & $X_d$ & 0 &
 0 & 0 & $\hat{X}_{\langle d,g\rangle}$ \\
 $\hat{X}_{f(t,p,q,v_p)}$ & $-\hat{X}_{f_t}$ & $\hat{X}_{\tilde{f}}$ &
 $- \hat{X}_{f_p}$ & $-\hat{X}_{\breve{f}}$ & 0 & 0 &
 $-\hat{X}_{\langle c,f\rangle}$ & $\hat{X}_{\ll f,g\gg}$
\end{tabular}}
\\ \rule[.5pt]{13.3cm}{1pt}
\end{table}
In table \thetable\ we present commutators of symmetry generators,
where the commutator $\left[X_i,X_j\right]$ is given at the
intersection of $i$th row and $j$th column. We have used here the
following shorthand notation:
\begin{eqnarray}
 & & c^\prime = tc_t + pc_p - c,\quad \tilde{g} = tg_t + pg_p - qg_q
  - v_pg_{v_p},\quad \hat{a} = ya_y + za_z - 2a \nonumber
      \\ & & \breve{g} = qg_q + v_pg_{v_p} - g,\quad \langle d,g\rangle =
  d_{tt}g_q + d_{tp}g_{v_p},\nonumber
 \\ & & \qquad\qquad\qquad\qquad\qquad \ll f,g\gg =   \frac{\partial(f,g)}{\partial(t,q)} +
  \frac{\partial(f,g)}{\partial(p,v_p)}.
  \label{notat}
\end{eqnarray}

\subsection{Action of recursion operator on symmetries\\ of Husain system}
 \label{sec-RHus}

Here we study the action of recursion operator (\ref{RHus}) on
symmetry characteristics of Husain system. We start with the
recursion
\begin{gather}
\left( \begin{array}{c}
    \tilde{\varphi}_1
    \\ \tilde{\psi}_1
 \end{array}
\right) =
  R\left(
 \begin{array}{c}
    \varphi_1
    \\ \psi_1
 \end{array}
\right) = R\left(
 \begin{array}{c}
    - q
    \\ q_yv_{pz} - q_zv_{py} + \varepsilon v_{pp}
 \end{array}
\right) = \varepsilon \left(
 \begin{array}{c}
    D_p^{-1}v_{pp}
    \\ q_p
 \end{array}
\right) \notag
 \\  = \varepsilon \left(
 \begin{array}{c}
    v_p
    \\ q_p
 \end{array}
\right) + \left(
 \begin{array}{c}
 c(y,z)\\
 0
 \end{array}
\right)
 = - \varepsilon \left(
 \begin{array}{c}
    \varphi_3
    \\ \psi_3
 \end{array}
\right) + \left(
 \begin{array}{c}
    \varphi_{5c(y,z)}
    \\ \psi_{5c(y,z)}
 \end{array}
\right),
 \label{RfiHus1}
\end{gather}
where the ``constant'' of integration $C(y,z,t) = c(y,z)$ is
time-independent due to the constraint $\tilde{\varphi}_{t} =
\tilde{\psi}$. We continue with the formula
\begin{gather}
\left( \begin{array}{c}
    \tilde{\varphi}_2
    \\ \tilde{\psi}_2
 \end{array}
\right) =
  R\left(
 \begin{array}{c}
    \varphi_2
    \\ \psi_2
 \end{array}
\right) = R\left(
 \begin{array}{c}
 t q + pv_p \\
 q + pq_p + t(q_zv_{py} - q_yv_{pz} - \varepsilon v_{pp})
 \end{array}
\right) \notag
 \\ \qquad\qquad\qquad\qquad\qquad = \left(
 \begin{array}{c}
    D_p^{-1}D_p\left[pq - \varepsilon tv_p\right]
    \\ p(q_zv_{py} - q_yv_{pz} - \varepsilon v_{pp}) - \varepsilon
    (tq_p + v_p)
 \end{array}
\right) \notag
 \\ \qquad\qquad\qquad = \left(
 \begin{array}{c}
 \varphi_{f = pq - \varepsilon tv_p} \\
 \psi_{f = pq - \varepsilon tv_p}
 \end{array}
\right) +  \left(
 \begin{array}{c}
    \varphi_{5c(y,z)}
    \\ \psi_{5c(y,z)}
 \end{array}
\right),
 \label{RfiHus2}
\end{gather}
where the first term in the right-hand side of the last equation
is a two-component characteristic of second-order symmetry
(\ref{gen2Hus}) with $f = pq - \varepsilon tv_p$, that obviously
satisfies conditions (\ref{eqf}). Now we compute
\begin{gather}
 \left( \begin{array}{c}
    \tilde{\varphi}_3
    \\ \tilde{\psi}_3
 \end{array}
\right) =
  R\left(
 \begin{array}{c}
    \varphi_3
    \\ \psi_3
 \end{array}
\right) = R\left(
 \begin{array}{c}
 - v_p \\ - q_p
 \end{array}
\right) = - \left(
 \begin{array}{c}
 q \\
 q_zv_{py} - q_yv_{pz} - \varepsilon v_{pp}
 \end{array}
\right)
  \notag
 \\ \qquad \qquad\qquad \qquad\qquad\;\; \mbox{} + \left(
 \begin{array}{c}
 c(y,z) \\ 0
 \end{array}
\right) = - \left(
 \begin{array}{c}
    \varphi_1
    \\ \psi_1
 \end{array}
\right) + \left(
 \begin{array}{c}
    \varphi_{5c(y,z)}
    \\ \psi_{5c(y,z)}
 \end{array}
\right).
 \label{RfiHus3}
\end{gather}
At the next step, we obtain the first nonlocal symmetry
\begin{gather}
  \left( \begin{array}{c}
    \tilde{\varphi}_4
    \\ \tilde{\psi}_4
 \end{array}
\right) =
  R\left(
 \begin{array}{c}
    \varphi_4
    \\ \psi_4
 \end{array}
\right) = R\left(
 \begin{array}{c}
v - \frac{1}{2}\,(yv_y + zv_z)
\\[2pt] q - \frac{1}{2}\,(yq_y + zq_z)
 \end{array}
\right) \notag
 \\ \mbox{} \qquad \qquad \qquad \qquad \quad \; = \frac{1}{2} \left(
 \begin{array}{c}
D_p^{-1} \left[v_{py}w_z - v_{pz}w_y - w_t \right] \\
q_yw_z - q_zw_y + \varepsilon w_p
 \end{array}
\right),
 \label{RfiHus4}
\end{gather}
where we have denoted $w = yv_y + zv_z - 2v$, so that $w_t = yq_y
+ zq_z - 2q$. Next, we obtain
\begin{gather}
  \left( \begin{array}{c}
    \tilde{\varphi}_{5a(y,z)}
    \\ \tilde{\psi}_{5a(y,z)}
 \end{array}
\right) =
  R\left(
 \begin{array}{c}
    \varphi_{5a(y,z)}
    \\ \psi_{5a(y,z)}
 \end{array}
\right) = R\left(
 \begin{array}{c}
 a(y,z) \\ 0
 \end{array}
\right) = \left(
 \begin{array}{c}
a_yv_z - a_zv_y \\
a_yq_z - a_zq_y
 \end{array}
\right) \notag
 \\ \qquad \qquad \qquad \qquad \quad \quad \mbox{} + \left(
 \begin{array}{c}
 c(y,z) \\ 0
 \end{array}
\right) = - \left( \begin{array}{c}
 \varphi_{6a(y,z)} \\ \psi_{6a(y,z)}
\end{array}
\right) + \left( \begin{array}{c}
 \varphi_{5c(y,z)} \\ \psi_{5c(y,z)}
\end{array}
\right) ,
 \label{RfiHus5}
\end{gather}
Another nonlocal symmetry stems out from the transformation of
symmetry $X_{6a(y,z)}$:
\begin{gather}
  \left( \begin{array}{c}
    \tilde{\varphi}_{6a(y,z)}
    \\ \tilde{\psi}_{6a(y,z)}
 \end{array}
\right) =
  R\left(
 \begin{array}{c}
    \varphi_{6a(y,z)}
    \\ \psi_{6a(y,z)}
 \end{array}
\right) = R\left(
\begin{array}{c}
  a_zv_y - a_yv_z \\ a_zq_y - a_yq_z
\end{array}
\right) \notag
 \\ \qquad \qquad \qquad \qquad \qquad  = \left(
 \begin{array}{c}
 D_p^{-1}\!\left[v_{pz}\varphi_{6a,y} - v_{py}\varphi_{6a,z} + \psi_{6a}\right]
 \\ q_z\varphi_{6a,y} - q_y\varphi_{6a,z} - \varepsilon \varphi_{6a,p}
 \end{array}
\right).
 \label{RfiHus6}
\end{gather}
Remaining point symmetry transforms under recursion in the
following way:
\begin{gather}
  \left( \begin{array}{c}
    \tilde{\varphi}_{c(t,p)}
    \\ \tilde{\psi}_{c(t,p)}
 \end{array}
\right) =
  R\left(
 \begin{array}{c}
    \varphi_{c(t,p)}
    \\ \psi_{c(t,p)}
 \end{array}
\right) = R \left(
 \begin{array}{c}
 c_t(t,p) \\ c_{tt}(t,p)
 \end{array}
\right) = \left(
 \begin{array}{c}
 D_p^{-1}c_{tt}(t,p) \\ - \varepsilon c_{tp}(t,p)
 \end{array}
\right)
 \notag
 \\[2pt] \qquad\qquad =  \left(
 \begin{array}{c}
 D_p^{-1}d_{tp}(t,p) \\ d_{tt}(t,p)
  \end{array}
\right)  =  \left(
 \begin{array}{c}
 d_t(t,p) \\ d_{tt}(t,p)
   \end{array}
\right) +  \left(
 \begin{array}{c}
 c_1(y,z) \\ 0
 \end{array}
 \right) \notag
 \\  \qquad\qquad\qquad\qquad\qquad\qquad\qquad\quad\; = \left(
 \begin{array}{c}
    \varphi_{d(t,p)}
    \\ \psi_{d(t,p)}
 \end{array}
\right) + \left( \begin{array}{c}
 \varphi_{5c_1(y,z)} \\ \psi_{5c_1(y,z)}
\end{array}
\right) ,
   \label{RfiHusc}
\end{gather}
where $c_1(y,z)$ is a ``constant'' of integration. Here $d(t,p)$
is related to $c(t,p)$ by the transformation $d_p = c_t$, $d_t = -
\varepsilon c_p$, so that $d(t,p)$ satisfies the same equation
(\ref{ceq}) as $c(t,p)$: $d_{tt} + \varepsilon d_{pp} = 0$. The
existence of the potential $d(t,p)$ for $c(t,p)$ follows from
(\ref{ceq}) presented in the form $(c_t)_t = (- \varepsilon
c_p)_p$.

Finally, we consider the action of $R$ on second-order symmetries
$X_f$ in (\ref{gen2Hus}) with $f = f(t,p,q,v_p)$ satisfying four
linear equations (\ref{eqf}):
\begin{gather}
   \left( \begin{array}{c}
    \tilde{\varphi}_{f(t,p,q,v_p)}
    \\ \tilde{\psi}_{f(t,p,q,v_p)}
 \end{array}
\right) =
  R\left(
 \begin{array}{c}
    \varphi_{f(t,p,q,v_p)}
    \\ \psi_{f(t,p,q,v_p)}
 \end{array}
\right) = R \left(
 \begin{array}{c}
 f(t,p,q,v_p) \\ f_t + q_tf_q + q_pf_{v_p}
 \end{array}
\right) \notag
 \\ \qquad \qquad \qquad \quad = \left(
 \begin{array}{c}
 D_p^{-1}\left(f_t - \varepsilon v_{pp}f_q +  q_pf_{v_p}\right)
 \\ - \varepsilon (f_p + q_pf_q) + q_tf_{v_p}
 \end{array}
\right),
 \label{RfiHus2f}
\end{gather}
where $q_t = q_zv_{py} - q_yv_{pz} - \varepsilon
 v_{pp}$. Equations (\ref{eqf}) imply the existence of a
potential $g(t,p,q,v_p)$ for $f$ defined by the relations
\begin{equation}\label{g_to_f}
  g_t = - \varepsilon f_p,\quad g_p = f_t,\quad g_q = f_{v_p},
  \quad g_{v_p} = - \varepsilon f_q.
\end{equation}
As a consequence of (\ref{g_to_f}), $g(t,p,q,v_p)$ satisfies the
same equations (\ref{eqf}) as $f(t,p,q,v_p)$. In terms of $g$, our
result (\ref{RfiHus2f}) becomes
\begin{gather}
   \left( \begin{array}{c}
    \tilde{\varphi}_f
    \\ \tilde{\psi}_f
 \end{array}
\right) =
  R\left(
 \begin{array}{c}
    \varphi_f
    \\ \psi_f
 \end{array}
\right) = \left(
 \begin{array}{c}
 D_p^{-1}\left(g_p + v_{pp}g_{v_p} +  q_pg_q\right)
 \\ g_t + q_pg_{v_p} + q_tg_q
 \end{array}
\right) = \left(
 \begin{array}{c}
 D_p^{-1}D_p[g] \\ D_t[g]
 \end{array}
\right) \notag
 \\  = \left(
 \begin{array}{c}
 g \\ D_t[g]
 \end{array}
\right) + \left(
 \begin{array}{c}
 c(y,z) \\ 0
 \end{array}
\right) = \left(
 \begin{array}{c}
    \varphi_{g(t,p,q,v_p)}
    \\ \psi_{g(t,p,q,v_p)}
 \end{array}
\right) + \left( \begin{array}{c}
 \varphi_{5c(y,z)} \\ \psi_{5c(y,z)}
\end{array}
\right),
 \label{RfiHus2fin}
\end{gather}
so that, up to an arbitrary ``constant'' of integration $c(y,z)$,
the recursion acts on the space of second-order symmetries and on
solutions of linear system (\ref{eqf}). We note that the second
application of this recursion takes us back to the original
second-order symmetry, up to the factor $- \varepsilon$:
\begin{equation}
   \left( \begin{array}{c}
    \tilde{\tilde{\varphi}}_f
    \\ \tilde{\tilde{\psi}}_f
 \end{array}
\right) =
  R\left(
 \begin{array}{c}
    \tilde{\varphi}_f
    \\ \tilde{\psi}_f
 \end{array}
\right) = - \varepsilon \left(
 \begin{array}{c}
    \varphi_f
    \\ \psi_f
 \end{array}
\right) - \left(
 \begin{array}{c}
    \varphi_{6c(y,z)}
    \\ \psi_{6c(y,z)}
 \end{array}
\right) + \left(
 \begin{array}{c}
    \varphi_{5d(y,z)}
    \\ \psi_{5d(y,z)}
 \end{array}
\right)
 \label{R2fiHus2}
\end{equation}
modulo ``constants'' of integration $c(y,z)$ and $d(y,z)$.

\subsection{Hamiltonian structure of symmetry flows and\\
conservation laws for Husain system}
 \label{sec-HamilHus}

 Presenting Lie equations for variational symmetries with the two-component
characteristics $(\varphi_i, \psi_i)$ in the Hamiltonian form
\begin{equation}
 \left(
 \begin{array}{c}
  v_{\tau_i} \\ q_{\tau_i}
 \end{array}
 \right) =
 \left(
 \begin{array}{c}
 \varphi_i \\ \psi_i
 \end{array}
 \right) =  J_0 \left(
 \begin{array}{c}
 \delta_v H^i \\ \delta_q H^i
 \end{array}
 \right),
 \label{Husflow}
 \end{equation}
we determine conserved densities $H^i$, corresponding to known
symmetry characteristics $(\varphi_i, \psi_i)$, using the inverse
Noether theorem
\begin{equation}
\left(
\begin{array}{c}
\delta_v H^i \\ \delta_q H^i
\end{array}
\right) = K \left(
\begin{array}{c}
\varphi_i \\ \psi_i
\end{array}
\right) = \left(
\begin{array}{c}
v_{py}\varphi_{iz} - v_{pz}\varphi_{iy} - \psi_i \\
\varphi_i
\end{array} \right),
 \label{HusNoether}
\end{equation}
where the symplectic operator $K = J_0^{-1}$ is defined in
(\ref{Khus}). Using this relation for symmetries from the list
(\ref{pointHus}), we find the corresponding Hamiltonians which
serve also as conserved densities for Husain system
\begin{eqnarray}
 & & X_1:\ H^1 = \frac{1}{2}\,\Big(\varepsilon v_p^2 - q^2\Big) = - H_1,
 \quad X_2:\ H^2 = p v_p q + \frac{1}{2}\,t\Big(q^2 - \varepsilon v_p^2 \Big),
 \nonumber
 \\ & & X_3:\ H^3 = - qv_p = - H_0,\quad
  X_5^{a(y,z)}:\ H^5_{a(y,z)} = a q + \frac{1}{2}\,\Big(a_yv_z -
 a_zv_y\Big)v_p ,
  \nonumber
 \\ & & X_6^{a(y,z)}:\ H^6_a = q(a_zv_y - a_yv_z) - v_yv_z (a_zv_{py} +
 a_yv_{pz}), \nonumber
 \\ & & X_c:\ H^c = c_t(t,p)q + \varepsilon
 c_{pp}(t,p)v ,
\label{HamHus}
\end{eqnarray}
where $c(t,p)$ satisfies Eq.~(\ref{ceq}). For the symmetry $X_4$,
Hamiltonian does not exist and so it is not a variational
symmetry.

Lie equations (\ref{order2Hus}) of second-order symmetries of
Husain system (\ref{2comp}) can also be presented in the
Hamiltonian form (\ref{Husflow}) with the Hamiltonian density
\begin{equation}\label{HamilHus2}
  H^f = F(t,p,q,v_p),
\end{equation}
where $F$ is defined in terms of the function $f(t,p,q,v_p)$
defined in (\ref{order2Hus}) as $F_q(t,p,q,v_p) = f$. On account
of equations (\ref{eqf}) and (\ref{HusNoether}) with $H=H^f$ and
$\varphi = \varphi_f$, $\psi = \psi_f$ defined by
(\ref{order2Hus}), $F$ can be shown to satisfy the equations
\begin{eqnarray}
 & & F_{tv_p} + \varepsilon F_{pq} = \alpha(t,p,v_p), \quad F_{pv_p} - F_{qt}
= 0, \nonumber \\
 & & F_{qq} + \varepsilon  F_{v_pv_p} = 0,
 \qquad \qquad F_{tt} + \varepsilon F_{pp} = \delta(t,p,v_p),
 \label{eqF}
\end{eqnarray}
where all the functions in the right-hand sides of
Eqs.~(\ref{eqF}) are arbitrary functions of their arguments. We
note, in particular, that if $H^f = F =\frac{1}{2}\,(q^2 -
\varepsilon v_p^2)$ in (\ref{HamilHus2}), then $F$ satisfies all
the conditions (\ref{eqF}) with $\alpha=0$, $\delta=0$ and
Hamiltonian (\ref{HamilHus2}) reduces to Hamiltonian (\ref{Hus1})
of Husain system, while second-order Lie equations
(\ref{order2Hus}) with $\tau = t$ coincide with the Husain system
(\ref{2comp}). Therefore, Husain system is embedded into the
hierarchy of second-order flows commuting with it.

Hamiltonians of the symmetry flows, presented here, serve as
conserved densities for Husain system. We note that the
conservation laws presented in this section seem to be different
from those given in the paper of V. Husain \cite{husain}.

\subsection{Action of recursion operator on Hamiltonians\\ of
symmetry flows}
 \label{sec-RH_Hus}

Transformation (\ref{RfiHus1}) of the symmetry
$(\varphi_1,\psi_1)$ corresponds to the following transformation
of Hamiltonian $H^1$
\begin{equation}\label{HusH1tr}
  \tilde{H}^1 = - \varepsilon H^3 + H^5_{c(y,z)} = q[\varepsilon v_p +
  c(y,z)] + \frac{1}{2}(c_yv_z - c_zv_y)v_p,
\end{equation}
where $c(y,z)$ is a ``constant'' of integration in
(\ref{RfiHus1}). For Hamiltonian $H_1$ of Husain system given in
(\ref{Hus1}), due to $H_1 = - H^1$, we have from (\ref{HusH1tr})
\begin{equation}\label{RfiHus_1}
 \tilde{H}_1 = \varepsilon H^3 - H^5_{c(y,z)} = - \varepsilon qv_p -
 H^5_{c(y,z)}= - \varepsilon H_0 - H^5_{c(y,z)},
\end{equation}
where $H_0$ is the second Hamiltonian (\ref{H0Hus}) in the
bi-Hamiltonian representation (\ref{biHamHus}) of Husain system,
that is, $\tilde{H}_1 = - \varepsilon H_0$ modulo arbitrary
``constant'' of integration $c(y,z)$.

Action of $R$ on the symmetry $(\varphi_2,\psi_2)$ in
(\ref{RfiHus2}) implies the following transformation of the
Hamiltonian $H^2$
\begin{equation}\label{HusH2tr}
  \tilde{H}^2 = H^{f=pq-\varepsilon tv_p} + H^5_{c(y,z)} =
  \frac{p}{2}\left(q^2 - \varepsilon v_p^2\right) - \varepsilon tqv_p  +
  H^5_{c(y,z)},
\end{equation}
that is, $H^2$ transforms to the Hamiltonian $H^f = F$ in
(\ref{HamilHus2}), where $F$ is determined by $f=pq-\varepsilon
tv_p$ due to the relation $F_q(t,p,q,v_p) = f$ and equations
(\ref{eqF}) for $F$. It is obvious that the specified $f$
satisfies all the equations (\ref{eqf}).

Transformation (\ref{RfiHus3}) of the symmetry $(\varphi_3,
\psi_3)$ results in the following transformation of Hamiltonian
$H^3$:
\begin{equation}\label{HusH3tr}
  \tilde{H}^3 = - H^1 + H^5_{c(y,z)} = \frac{1}{2}(\varepsilon v_p^2 - q^2) +
  H^5_{c(y,z)}.
\end{equation}
Transformation (\ref{RfiHus5}) of the symmetry $(\varphi_5,
\psi_5)$ implies
\begin{equation}\label{HusH5tr}
  \tilde{H}^5_{a(y,z)} = - H^6_{a(y,z)} +
  H^5_{c(y,z)} = q(a_yv_z - a_zv_y) + v_yv_z(a_zv_{py} +
  a_yv_{pz}) + H^5_{c(y,z)}.
\end{equation}
Transformation of the symmetry $(\varphi_6, \psi_6)$ gives rise to
nonlocal flow (\ref{RfiHus6}) and hence Hamiltonian
$\tilde{H}^6_{a(y,z)}$ is also nonlocal with respect to the
Hamiltonian operator $J_0$. In the next section, in
Eq.~(\ref{RHus6}) we will see that the Hamiltonian of this
nonlocal flow will be a local one, notably $H^6_{a(y,z)}$, if
considered with respect to second Hamiltonian operator $J_1$.

From transformation (\ref{RfiHusc}) of the symmetry
$(\varphi_{c(t,p)}, \psi_{c(t,p)})$ we deduce the transformation
of the Hamiltonian $H^{c(t,p)}$
\begin{equation}\label{HusHctr}
  \tilde{H}^c = H^d + H^5_{c(y,z)} = q\left[d_t(t,p) + c(y,z)\right] +
  \varepsilon d_{pp}(t,p)v + \frac{1}{2}(c_yv_z - c_zv_y)v_p ,
\end{equation}
where $d(t,p)$ is related to $c(t,p)$ by the equations $d_p =
c_t$, $d_t = - \varepsilon c_p$ and satisfies the same equation
(\ref{ceq}) as $c(t,p)$: $d_{tt} + \varepsilon d_{pp} = 0$.

Finally, transformation (\ref{RfiHus2fin}) of the general
second-order symmetry $(\varphi_f, \psi_f)$, with $f(t,p,q,v_p)$
satisfying equations (\ref{eqf}), results in the following
transformation of the Hamiltonian $H^f$:
\begin{equation}\label{HusHftr}
  \tilde{H}^f = H^g + H^5_{c(y,z)} = G(t,p,q,v_p) + c(y,z)q +
  \frac{1}{2}(c_yv_z - c_zv_y)v_p ,
\end{equation}
where $G_q = g$ and $g(t,p,q,v_p)$ is determined by equations
(\ref{g_to_f}) for any given $f(t,p,q,v_p)$ satisfying
(\ref{eqf}). From (\ref{R2fiHus2}) we note that the repeated
application of the recursion to $\tilde{H}^f$ takes us back to
$H^f$: $\tilde{\tilde{H}}^f = - \varepsilon H^f$ modulo
``constants'' of integration. Since the Hamiltonian $H_1$ of
Husain system is a particular case of $H^f$ with $f=q$, the same
is true for $H_1$: $\tilde{\tilde{H}}_1 =  - \varepsilon H_1$.
Similarly, the second Hamiltonian $H_0$ of Husain system can be
obtained from $H^f$ at $f = v_p$, which yields $\tilde{H}_0 = H_1$
and $\tilde{\tilde{H}}_0 = - \varepsilon H_0$.

\section{Hierarchy and bi-Hamiltonian representations for symmetry flows
of Husain system}
 \setcounter{equation}{0}
 \label{sec-highHus}

Similar to the beginning of section \ref{sec-highflows}, in this
section, in contrast to subsections \ref{sec-RHus} and
\ref{sec-RH_Hus}, we have $D_p^{-1} =
\int_{-\infty}^p\,dp^{\,\prime}$ (for functions vanishing rapidly
at $-\infty$) in the definitions (\ref{RHus}) and (\ref{J1Hus}) of
$R$ and $J_1$, respectively, so that $D_p^{-1}D_p = 1$. In section
\ref{sec-highflows}, we have noted the identity $RJ_0 = J_1 =
J_0R^\dag$ resulting in the relation (\ref{R+delH}), which
signifies that the action of $J_1$ on variational derivatives of
the Hamiltonian $H$ can be replaced by the action of $J_0$ on
variational derivatives of the Hamiltonian $\tilde{H}$ obtained
from $H$ by the action of $R$, in accordance with the formulas
derived in subsection \ref{sec-RH_Hus}, where we now skip all the
terms involving arbitrary ``constants'' of integration.

We now proceed to use relation (\ref{R+delH}) for constructing
hierarchies of Husain system and its symmetry flows, together with
bi-Hamiltonian representations of the symmetry flows. Applying $R$
to Husain system in Hamiltonian form (\ref{HamilHus}), we obtain
\begin{equation}\label{hi1HamHus}
\left(
\begin{array}{c}
v_{t_1} \\ q_{t_1}
\end{array}
\right) =  J_1 \left(
\begin{array}{c}
\delta_v H_1 \\ \delta_q H_1
\end{array}
\right) = J_0 \left(
\begin{array}{c}
\delta_v \tilde{H}_1 \\ \delta_q \tilde{H}_1
\end{array}
\right)
 = - \varepsilon J_0 \left(
\begin{array}{c}
\delta_v H_0 \\ \delta_q H_0
\end{array}
\right)
 = - \varepsilon \left(
\begin{array}{c}
 v_p \\ q_p
\end{array}
\right),
\end{equation}
where $t_1 = \tilde{t}$ is the parameter of the group transformed
by $R$ and we have used $\tilde{H}_1 = - \varepsilon H_0$ due to
(\ref{RfiHus_1}) modulo $H^5_{c(y,z)}$. The second application of
$R$ to Hamiltonian system (\ref{hi1HamHus}) yields
\begin{gather}
\left(
\begin{array}{c}
v_{t_2} \\ q_{t_2}
\end{array}
\right) =  J_1 \left(
\begin{array}{c}
\delta_v \tilde{H}_1 \\ \delta_q \tilde{H}_1
\end{array}
\right) = J_0 \left(
\begin{array}{c}
\delta_v \tilde{\tilde{H}}_1 \\ \delta_q \tilde{\tilde{H}}_1
\end{array}
\right) = - \varepsilon J_0 \left(
\begin{array}{c}
\delta_v H_1 \\ \delta_q H_1
\end{array}
\right) \notag
 \\ \qquad \qquad \qquad \qquad \qquad \quad = - \varepsilon \left(
\begin{array}{c}
 q \\ q_zv_{py} - q_yv_{pz} - \varepsilon v_{pp}
\end{array}
\right)
 \label{hi2HamHus}
\end{gather}
due to $\tilde{\tilde{H}}_1 =  - \varepsilon H_1$, so that we are
back to Husain system and further applications of the recursion
operator will not generate an infinite hierarchy. We also note
that applying $J_0$ to $H_0$ will not yield anything new
\begin{equation}\label{hi0HamHus}
\left(
\begin{array}{c}
v_{t_0} \\ q_{t_0}
\end{array}
\right) =  J_0 \left(
\begin{array}{c}
\delta_v H_0 \\ \delta_q H_0
\end{array}
\right) = \left(
\begin{array}{c}
 v_p \\ q_p
\end{array}
\right),
\end{equation}
while $J_1$ applied to $H_0$ is equivalent to $J_0$ applied to
$\tilde{H}_0 = H_1$, which again results in bi-Hamiltonian
representation (\ref{biHamHus}) of Husain system.

Next we apply $R$ to Hamiltonian symmetry flows generated by
(\ref{pointHus}) that commute with Husain flow (\ref{2comp}). We
skip the flow of $H^1 = - H_1$, which will reproduce
(\ref{hi1HamHus}) up to a sign, and start with $H^2$ from
(\ref{HamHus}) with the following results
\begin{eqnarray}\label{RHus2}
 & & \left(
  \begin{array}{c}
 v_{\tilde{t}^2} \\ q_{\tilde{t}^2}
  \end{array}
  \right)
  = J_1 \left(
  \begin{array}{c}
  \delta_vH^2 \\ \delta_qH^2
  \end{array}
  \right) = J_0 \left(
  \begin{array}{c}
  \delta_v\tilde{H}^2 \\ \delta_q\tilde{H}^2
  \end{array}
  \right)\nonumber
  \\ & & \qquad\qquad = \left(
  \begin{array}{c}
 pq - \varepsilon tv_p \\ p(v_{py}q_z - v_{pz}q_y) - \varepsilon
 (pv_p + tq)_p
 \end{array}
  \right) ,
\end{eqnarray}
where $\tilde{H}^2 = \frac{\displaystyle p}{\displaystyle
2}\left(q^2 - \varepsilon v_p^2\right) - \varepsilon tqv_p = pH_1
- \varepsilon tH_0$ according to (\ref{HusH2tr}).

Since $H^3 = - H_0$, we skip the action of $J_1$ on $H^3$. There
is no Hamiltonian for the symmetry generated by $X_4$ in
(\ref{pointHus}). Therefore, we proceed with the action of $J_1$
on $H^5_a(y,z)$ in (\ref{HamHus}) to obtain
\begin{gather}
  \left(
\begin{array}{c}
v_{\tilde{t}^5} \\ q_{\tilde{t}^5}
\end{array}
\right)
  =  J_1 \left(
\begin{array}{c}
\delta_v H^5_{a(y,z)} \\ \delta_q H^5_{a(y,z)}
\end{array}
\right) = J_0  \left(
 \begin{array}{c}
 \delta_v \tilde{H}^5_a \\ \delta_q \tilde{H}^5_a
 \end{array}  \right)
 = - J_0 \left(
\begin{array}{c}
\delta_v H^6_a \\ \delta_q H^6_a
\end{array}
\right) = - \left(
\begin{array}{c}
 \varphi_{6a} \\ \psi_{6a}
\end{array}  \right)
 \notag
 \\  \qquad \qquad \qquad \qquad \qquad \qquad \qquad  = \left(
  \begin{array}{c}
   a_yv_z - a_zv_y \\[1mm] a_yq_z - a_zq_y
  \end{array}  \right) ,
  \label{RHus5}
\end{gather}
where we have used that $\tilde{H}^5_a = - H^6_a$ according to
(\ref{HusH5tr}).

Applying $J_1$ to the flow of the symmetry $(\varphi_{6a},
\psi_{6a})$, defined in (\ref{pointHus}), with the Hamiltonian
$H^6_{a(y,z)}$ given in (\ref{HamHus}), we obtain the first
nonlocal symmetry flow (\ref{RfiHus6}) in the hierarchy of
variational symmetries of Husain system
\begin{eqnarray}
 & & \left(
\begin{array}{c}
v_{\tilde{t}^6} \\ q_{\tilde{t}^6}
\end{array}
\right) = J_1  \left(
 \begin{array}{c}
 \delta_v H^6_a \\[1pt] \delta_q H^6_a
 \end{array}  \right)
  =  J_0 \left(
\begin{array}{c}
\delta_v \tilde{H}^6_a \\[1pt] \delta_q \tilde{H}^6_a
\end{array}
\right) \nonumber
 \\  & & \qquad\qquad = \left(
 \begin{array}{c}
 D_p^{-1}\!\left[v_{pz}\varphi_{6a,y} - v_{py}\varphi_{6a,z} + \psi_{6a}\right]
 \\ q_z\varphi_{6a,y} - q_y\varphi_{6a,z} - \varepsilon \varphi_{6a,p}
 \end{array}
\right),
 \label{RHus6}
\end{eqnarray}
because $D_p^{-1}$ obviously acts not on a total $p$-derivative.
Since $X_5^a$ and $X_6^a$ commute, considerations presented at the
end of section \ref{sec-highflows} show that the hierarchy of
symmetry flows, generated by powers of $R$ from the symmetry
$X_5^a$, consists of commuting flows.

We skip a discussion of point symmetry $X_{c(t,p)}$ in
(\ref{pointHus}), since it is a particular case of the
second-order symmetry $X_{f(t,p,q,v_p)}$ in (\ref{gen2Hus}). So we
finally consider the action of $R$ on the Hamiltonian $H^f$,
defined in (\ref{HamilHus2})
\begin{gather}
  \left(
\begin{array}{c}
v_{\tilde{t}} \\ q_{\tilde{t}}
\end{array}
\right) = J_1  \left(
 \begin{array}{c}
 \delta_v H^f \\ \delta_q H^f
 \end{array}  \right)
  =  J_0 \left(
\begin{array}{c}
\delta_v \tilde{H}^f \\ \delta_q \tilde{H}^f
\end{array}
\right)
  =  J_0 \left(
\begin{array}{c}
\delta_v H^g \\ \delta_q H^g
\end{array}
\right) \notag
 \\ \qquad \qquad = \left(
  \begin{array}{c}
    g(t,p,q,v_p) \\[2pt] g_t + g_q(q_zv_{py} - q_yv_{pz}
- \varepsilon v_{pp}) + q_pg_{v_p}
  \end{array}  \right) ,
  \label{RHusf}
\end{gather}
where, according to (\ref{HusHftr}), $\tilde{H}^f = H^g$ with
$g(t,p,q,v_p)$ related to $f$ by equations (\ref{g_to_f}). Since
$\tilde{\tilde{H}}^f = - \varepsilon H^f$, the second application
of $R$ to $\tilde{H}^f$ takes us back to the original second-order
flow (\ref{gen2Hus}) with the generating function $f$. We note
that the flow (\ref{gen2Hus}) is a natural generalization of the
Husain system.

\section{Conclusion}

The importance of equations, that admit partner symmetries is that
they possess nonlocal recursion relations for symmetries of a
special form, that enable us to obtain noninvariant solutions by
symmetry methods. They are integrable equations also in a more
traditional sense because they admit Lax representation together
with infinite sets of symmetries and conservation laws. Mixed
heavenly equation, that combines, up to a point, first and second
heavenly equations of Pleba\~nski, and Husain heavenly equation
are among the simplest canonical equations with these properties.
We have reformulated both equations as two-component evolution
systems, which resulted in a natural definition of a single matrix
recursion operator for each system. This operator together with
the operator of the symmetry condition forms a Lax pair of
Olver-Ibragimov-Shabat type for each of these systems.

By choosing an appropriate Lagrangian, we have discovered
symplectic and Hamiltonian representations for both mixed heavenly
system and Husain system. Applying the recursion operator to the
Hamiltonian operator, we have explicitly generated second
Hamiltonian structures for these systems. Thus, we have shown that
the mixed heavenly equation and Husain equation, set in a
two-component form, are bi-Hamiltonian systems with compatible
Hamiltonian structures forming a Poisson pencil. Therefore, they
are integrable Hamiltonian systems also in the sense of Magri.
Hamiltonian structure relates symmetries and conserved densities
which serve as Hamiltonian densities for symmetry flows. We have
determined such Hamiltonian densities for all variational point
symmetries and generalized second-order symmetries for both mixed
heavenly system and Husain system. We obtained transformations of
symmetries and their Hamiltonians under the action of recursion.

We studied hierarchies of the mixed heavenly system and Husain
system. A characteristic feature of these systems is that the
repeated action of the recursion operator on each of these systems
takes us back to the original system, so that there are only two
members in the hierarchy containing each of these systems and no
infinite hierarchy can be generated from them by the recursion.
This is a remarkable distinctive feature of the mixed heavenly
system and Husain system as compared to the second heavenly
equation of Pleba\~nski and complex Monge-Amp\`ere equation whose
bi-Hamiltonian structures we studied earlier. However, we
discovered an infinite hierarchy of Hamiltonian flows commuting
with each other and with the considered system. Such a hierarchy
is generated by one of Lie point symmetries of each system. Higher
flows in each hierarchy are nonlocal and we have obtained
explicitly first nonlocal Hamiltonian flows for both systems.
Thus, the mixed heavenly system and Husain system possess an
infinite number of (mostly nonlocal) symmetries and conservation
laws which is a customary property of integrable systems. We have
obtained bi-Hamiltonian representations for the flows of all
variational point symmetries and all higher second-order
symmetries. Further study of higher-order and nonlocal symmetries
may provide an additional interesting information about the
structure of these hierarchies.

There is also a set of all higher second-order symmetries for each
system, which have a functional arbitrariness and commute with
this system but not with each other. The commutators of all the
symmetries are presented in the tables. This set of second-order
symmetries of each system includes the original system as a very
particular simple case and looks like a natural generalization of
mixed heavenly and Husain systems. It would be interesting to
study symmetries and conservation laws of these more general
systems of second-order Lie equations (\ref{Lieord2}) and
(\ref{order2Hus}). An important question is if they possess
partner symmetries, may be under some restrictions.

\section*{Acknowledgements}

The authors are grateful to A. V. Mikhailov for an interesting
discussion. The research of MBS was partly supported by the
research grant from Bogazici University Scientific Research Fund,
research project No. 07B301.

\end{document}